\documentclass[12pt,a4]{article} 

\usepackage{mathptmx}
\usepackage{helvet}
\usepackage{courier}
\usepackage{type1cm}
\usepackage{makeidx}         
\usepackage{graphicx}        
\usepackage{multicol}        
\usepackage[bottom]{footmisc}
\usepackage{color}

\makeindex             

\graphicspath{{./figures/}}
\usepackage{amsmath}
\usepackage{amssymb}
\usepackage{latexsym}
\usepackage{authblk}
\usepackage{geometry}
\usepackage{cite}
\usepackage{hyperref}
\begin{document}
\author{\normalsize  Dirk Helbing$^{1,2}$\thanks{Corresponding author.}~, Dirk Brockmann$^{3,4}$, Thomas Chadefaux$^{1}$, Karsten Donnay$^{1}$, Ulf Blanke$^{5}$, Olivia Woolley-Meza$^{1}$, Mehdi Moussaid$^{6}$, Anders Johansson$^{7,8}$, Jens Krause$^{9}$, Sebastian Schutte$^{10}$, and Matja{\v z} Perc$^{11}$\\
{\footnotesize
$^1$ Chair of Sociology, in particular of Modeling and Simulation, ETH Zurich, Swiss Federal Institute of Technology, 8092 Zurich, Switzerland\\
$^2$ Risk Center, ETH Zurich, Swiss Federal Institute of Technology, 8092 Zurich, Switzerland\\
$^3$ Robert Koch-Institute, 13353 Berlin, Germany\\
$^4$ Institute for Theoretical Biology, Humboldt-University, 10115 Berlin, Germany\\
$^5$ Wearable Computing Lab, ETH Zurich, Swiss Federal Institute of Technology, 8092 Zurich, Switzerland\\
$^6$ Center for Adaptive Rationality (ARC), Max Planck Institute for Human Development, 14195 Berlin, Germany\\
$^7$ Centre for Advanced Spatial Analysis, University College London, W1T 4TJ, London, UK\\
$^8$ Systems Centre, Department of Civil Engineering, University of Bristol, BS8 1UB, Bristol, UK\\
$^9$ Department of Biology and Ecology of Fishes, Leibniz-Institute of Freshwater Ecology and Inland Fisheries, 12587 Berlin, Germany\\
$^{10}$ Center for Comparative and International Studies, ETH Zurich, Swiss Federal Institute of Technology, 8092 Zurich, Switzerland\\
$^{11}$ Faculty of Natural Sciences and Mathematics, University of Maribor, SI-2000 Maribor, Slovenia\\
}
}

\title{\Large\bf Saving Human Lives: What Complexity Science and Information Systems can Contribute}
\maketitle
\vspace{-0.5cm}
\abstract{We discuss models and data of crowd disasters, crime, terrorism, war and disease spreading to show that conventional recipes, such as deterrence strategies, are often not effective and sufficient to contain them. Many common approaches do not provide a good picture of the actual system behavior, because they neglect feedback loops, instabilities and cascade effects. The complex and often counter-intuitive behavior of social systems and their macro-level collective dynamics can be better understood by means of complexity science. We highlight that a suitable system design and management can help to stop undesirable cascade effects and to enable favorable kinds of self-organization in the system. In such a way, complexity science can help to save human lives.
}
\clearpage

\section{Introduction}
Over the past decades complexity science has evolved from purely theoretical contributions towards applications with real-life relevance. In this paper, we will pay particular attention to the problem of collective dynamics in human populations~\cite{Castellano_2009, Helbing_Sociodynamics, Weidlich_2000}. Topics addressed range from crowd disasters~\cite{Helbing_2010} to crime~\cite{short_pnas10, perc2013und}, terrorism~\cite{Clauset_2007, Bohorquez_2009, Johnson_2011}, wars~\cite{Science_SI_2012, BecKinZen00, chadefaux2014}, and the epidemic spreading of diseases~\cite{Anderson_1992, Brockmann_2013}. These fields increasingly include quantitative, realistic, and even predictive models, bringing together statistical data analysis, modeling efforts, analytical approaches, and laboratory experiments. These advances make the research described in this paper increasingly relevant for real-life situations, creating the opportunity to employ scientific knowledge to save human lives.
\par
The diverse problems we consider exhibit systemic instabilities in which control of the macro-level, the collective dynamics, is lost even though there seems to be a reasonable level of control at the micro-level of individual system components~\cite{NATURE}. In traffic flows, for example, researchers have found that ``phantom'' traffic jams can occur in the absence of accidents or bottlenecks~\cite{Kerner_1993, Sugiyama_2008}. Despite the drivers' best efforts to prevent traffic jams, they are unavoidable if the vehicle density crosses a certain threshold~\cite{traf1, RMP}. At high densities, small variations in speed are amplified from one driver to the next, causing a cascade effect. This ends in a situation undesirable for all---one in which every driver is stopped. What, then, can be done to prevent these systemic instabilities? We suggest here that a better scientific understanding, based on methods from complexity science, can help. For example, despite the complexity of traffic patterns~\cite{Schoenhof_2007}, they can be described by analytic and predictive models~\cite{traf1}. This has enabled traffic assistance systems based on a distributed control approach and bottom-up self-organization, which can be used for effective congestion avoidance \cite{ACC,Stefan}. Surprisingly, well-designed self-regulation in this case outperforms classical top down control.
\par
The focus of this paper will be to give an overview of the new scientific understanding of crowd disasters, crime, terrorism, wars, and epidemics gained through a complex systems perspective. We will explain why ``linear thinking'' and classical control approaches fail to overcome or mitigate such problems efficiently due to the non-linear, non-equilibrium and therefore often counter-intuitive nature of these problems. Furthermore, we outline how complexity science can provide better solutions to some long-standing problems than conventional approaches. In particular, we will illustrate how this can contribute to addressing a number of serious issues still plaguing society despite better-than-ever science, measurement opportunities, information systems, and technology.

\section{Crowd disasters and how to avoid them}
\begin{figure}
\centering
\includegraphics[width=0.7\columnwidth,clip]{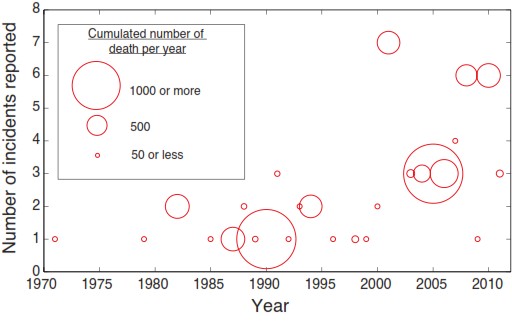}
\caption{Cumulative number of fatalities for major crowd disasters between 1970 and 2012. The figure shows a general upward trend in both the number of crowd disasters and the fatalities caused by them.}
\label{fig:disasters}
\end{figure}

Tragic crowd disasters repeatedly occur despite strict codes and guidelines for the organization of mass events. In fact, the number of crowd disasters and the overall number of fatalities are on the rise, probably as a result of the increasing frequency and size of mass events (see Fig.\ref{fig:disasters}). The classical approach to the avoidance of crowd disasters is reflected in the concept of ``crowd control'', which assumes that it is necessary to ``make people behave''. The concept bears some resemblance to how police sometimes handle violent demonstrations and riots. For example, tear gas and police cordons are used with the intention to gain control over crowds. However, there is evidence that such measures, which are usually intended to improve the safety of the crowd, may also unintentionally deteriorate a situation. For example, the use of tear gas may have played a significant role in the crowd disasters in Lima, Peru (1998), Durban, South Africa (2000), Lisbon, Portugal (2000), Harare, Zimbabwe (2000), and Chicago, Illinois (2003)~\cite{Helbing_2002}.

Our approach in the following is to clarify the cause of crowd disasters and then to discuss simple rules to improve crowd safety, oriented at giving more control to the individuals (``empowerment''). This strategy can be best described as creating a system design (``institutional setting'') enabling all individuals and stakeholders to make a beneficial contribution to the proper functioning of the system---here mass events.

\subsection{Crowd turbulence}
Crowd disasters are examples of situations in which people are killed by other people, even though typically nobody wants to harm anybody. To explain the reasons for crowd disasters, concepts such as ``pushing'', ``mass panic'', ``stampede'', ``crowd crushes'' or ``trampling'' have often been used. However, such views tend to blame the visitors of mass events, thereby preventing the proper understanding of crowd disasters and their avoidance in the future. In fact, social order will generally not break down during crowd disasters~\cite{johnson_1987} and it has been found that extreme events may lead to an emergent collective identity that will result in increased solidarity with strangers~\cite{drury} rather than the opposite.

\begin{enumerate}
\item The term ``pushing'' suggests that people are relentlessly pushing towards their destination, disregarding the situation of others. However, when the density in the crowd is very high, even inadvertent body movements will cause physical interactions with others. The forces transmitted by such involuntary body interactions may add up from one body to the next, thereby causing a situation where people are unintentionally pushed around in the crowd. This situation is hard to distinguish from intentional pushing.
\item Fatalities during a crowd disaster are often perceived as the result of a ``crowd crush'', i.e. a situation in which the pressure on human bodies becomes so high that it causes deadly injuries. Such a situation may arise in the case of a ``stampede'', i.e. people desperately rushing into one direction but then facing a bottleneck, such that the available space per person is increasingly reduced, until the density in the crowd becomes life-threatening. For example, the crowd disaster in Minsk, Belarus (1999), was caused by people fleeing into a subway station from heavy rain.
\item Another explanation of crowd disasters assumes a ``state of (psychological) panic'', which makes individuals behave in an irrational and relentless way, such that people are killed. An example of this kind may be the crowd disaster in Baghdad, Iraq (2005), which was triggered by rumors regarding an imminent suicide bomber. The assumption of mass panic interprets fatalities in crowd disasters similar to manslaughter by a rioting mob---an interpretation, which may warrant the use of tear gas or police cordons to control an ``outraged crowd''.
\end{enumerate}
In sharp contrast to this, it turns out that many crowd disasters do not result from a ``stampede'' or ``mass panic''~\cite{Loveparade}. Prior to 2007, crowd-flow theory generally assumed a state of equilibrium in the crowd, where each level of crowd density (1/m$^2$) could be mapped to a corresponding value of the crowd flow (1/m/s). This assumption turned out to be approximately correct for low-to-medium levels of crowd density. However, it was later found~\cite{Helbing_2007} that at very high density the crowd is driven far from equilibrium.
\par
In many cases, fatalities result from a phenomenon called ``crowd turbulence'', which occurs when the density in the crowd is so high that bodies automatically touch each other. As a result, physical forces are transmitted from one body to another---a process that is unavoidable under such conditions. These forces may add up and create force chains that cause sudden subsequent pushes from various directions, which can usually not be anticipated. Eventually, the individuals are pushed so much that one or some of them may stumble and fall to the ground. This creates a ``hole'' in the crowd, which breaks the balance of forces for the surrounding people: they are pushed from behind, but not anymore from the front, such that they are either forced to step on fallen persons (``trampling'') or they will also fall. Due to this domino effect, such ``holes'' eventually grow bigger and bigger (a situation that may be coined ``black hole effect''). This explains the phenomenon of piles of people, which are often identified as the causes of crowd disasters and interpreted as instances of trampling. Under such conditions, it is extremely difficult for individuals to get back on their feet and people on the ground eventually die of suffocation.
\par
The occurrence of ``crowd turbulence'' can be understood and reproduced in computer simulations by means of force-based models. Generally speaking, the behavior of a pedestrian results from two distinct kinds of mechanisms:
\begin{enumerate}
  \item The \textit{cognitive processes}~\cite{Moussaid_2011} used by a pedestrian during interactions with other individuals, which apply when people undertake avoidance manoeuvres, choose an avoidance side, plan a way to their destination, or coordinate their movements with others.
  \item The \textit{physical pressures} resulting from body contacts with neighboring individuals, which apply mostly in situations of overcrowding and are responsible for the phenomenon of crowd turbulence.
\end{enumerate}
 While the first category of interactions can be successfully described by means of a variety of methods (e.g. social forces, cognitive heuristics, cellular automata), the second type of interactions necessarily calls for force-based models, as it describes the result of unintentional movements due to physical pressures exerted by densely packed bodies.
\par
Therefore, crowd turbulence can be described by means of a contact force $\vec{f}_{ij}$ exerted by a pedestrian $j$ on another pedestrian $i$ defined as:
\begin{equation}
\vec{f}_{ij}=kg(r_{i}+r_{j}-d_{ij})\vec{n}_{ij}~.
\end{equation}

Herein, $\vec{n}_{ij}$ is the normalized vector pointing from pedestrian $j$ to $i$, and $d_{ij}$ is the distance between the pedestrians' centres of mass~\cite{Helbing_2000}. The parameter $k$ indicates the strength of between-body repulsion force. The function $g(x)$ is defined as
\begin{equation}
 g(x)=\begin{cases}
 x & \text{if pedestrians $i$ and $j$ are touching each other,}\\
 0 & \text{otherwise.}
\end{cases}
\end{equation}
For simplicity, the projection of the pedestrian's body on the horizontal plane is represented by a circle of radius $r_i = m_i/320$, where $m_i$ represents the mass of pedestrian $i$. The physical interaction with a wall $W$ is represented analogously by another contact force:
\begin{equation}
\vec{f}_{iW}=kg(r_{i}-d_{iW})\vec{n}_{iW}~.
\end{equation}
Here $d_{iW}$ is the distance to the wall $W$ and $\vec{n}_{iW}$ is the direction perpendicular to it. Therefore, both contact forces $\vec{f}_{ij}$ and $\vec{f}_{iW}$ are nonzero only when the pedestrian $i$ is in physical contact with a wall or another individual. Under normal walking conditions (that is, when no body contacts occur), the movement of pedestrians is determined by strategic and cognitive processes that can be described by means of several distinct methods such as social forces~\cite{Helbing_1995} or cognitive heuristics~\cite{Moussaid_2011}. Nevertheless, the method that is used to describe the free movements of pedestrians has little effect on the dynamics that emerge at extreme densities, since pedestrians movements are mostly unintentional in situations of overcrowding. Describing such situations we may thus use a generic component $\vec{f}_{i}^{0}$ describing how the pedestrian $i$ would move under normal walking conditions. The resulting acceleration equation in this case then reads
\begin{equation}
d\vec{v}_{i}/dt=\vec{f}_{i}^{0}+\sum_{j}\vec{f}_{ij} / m_i+ \sum_{W}\vec{f}_{iW} / m_i~,
\end{equation}
where $\vec{v}_{i}$ denotes the speed of pedestrian $i$, where the component $\vec{f}_{i}^{0}$ is negligible under extremely crowded conditions. The acceleration equation can be solved together with the usual equation of motion
\begin{equation}
d\vec{x}_{i}/dt=\vec{v}_{i}~,
\end{equation}
where $\vec{x}_{i}$ denotes the location of pedestrian $i$ at time $t$.
\par
\begin{figure}
\centering
\includegraphics[width=0.4\columnwidth]{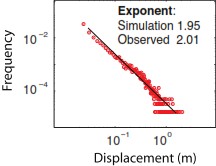}
\caption{Result of computer simulations of the above model indicating a power law distribution of displacements at extreme density. The displacement of a pedestrian $i$ is the change of location of that individual between two subsequent stops where the speed of $i$ is less than 0.05 m/s. Simulations were conducted at a density of 6 pedestrians per $m^2$, corresponding to a unidirectional flow of 360 pedestrians in a corridor of length 10m and width 6m with a bottleneck of width 4m. Model parameters were set to $k=5\cdot10^3$ and the body masses $m_i$ were uniformly distributed in the interval [60kg,100kg].}
\label{fig:powerLaw}
\end{figure}

Computer simulations of the above model in crowded situations, where physical interactions dominate over intentional movements, give rise to global breakdowns of coordination, where strongly fluctuating and uncontrollable patterns of motion occur. Crowd turbulence is particularly likely around bottlenecks, where local increases of the pedestrian density enhance the propagation of physical pressures from one individual to another. In particular, the unbalanced pressure distribution results in sudden stress releases and earthquake-like mass displacements of many pedestrians in all possible directions, which is well approximated by a power law with an exponent of $1.95$ (see Fig.~\ref{fig:powerLaw}). This result is in good agreement with empirical observations of crowd turbulence, which exhibit a power law distribution with exponent $2.01$~\cite{Helbing_2007}.

\par

\subsection{Simple rules for safety}
\label{sec:rulessafety}
Preparing for mass events is a major challenge, as there is a host of things that can go wrong over the course of an event (for a recent example see Ref.~\cite{Loveparade}). Therefore, the overall strategy is to develop and implement a suitable system design, which can be well operated under normal, but also challenging conditions. This includes institutional settings (e.g. spatial boundary conditions) and the way interactions between different stakeholders are organized. All elements of the plan, should be exercised before, including contingency plans. In case of the reorganization of the annual pilgrimage in Saudi-Arabia (``Hajj''), for example, many elements were combined to increase the level of safety. This included~\cite{Crowdsafety}:
\begin{enumerate}
\item advance information,
\item good signage,
\item reliable communication,
\item a control tower jointly used by all responsible authorities,
\item a new Jamarat Bridge design (with separate ramps for pilgrims coming from and going to different areas),
\item a unidirectional flow organization,
\item a combined routing concept and scheduling program,
\item a real-time flow monitoring,
\item a re-routing in  situations of high capacity utilization of certain routes,
\item contingency plans for all kinds of situations (including bad weather conditions).
\end{enumerate}
It is important to consider that the organization of a mass event must also always take into account the natural (physical, physiological, psychological, social and economic) needs of humans, such as sufficient space, water, food, toilet facilities, perceived progress towards the goal, feeling of safety, information, communication, etc. Neglecting such factors can promote crowd disasters, in particular, if several shortcomings come together.
\par
While sophisticated procedures combining expert consulting, computer simulations and video monitoring remain the safest way to avoid accidents, most event organizers are not in a position to undertake them all. In fact, such complex procedures are often expensive and turn out to be cost-effective only for large events, or if mass events are regularly held at the same place (e.g. the Muslim pilgrimage in Mecca). Hence, organizers of small- and intermediate-size events are often constrained to rely on their intuition and past experience  to assess the safety of the events they are in charge of. However, the dynamics of crowd behavior is complex and often counter-intuitive. A systemic failure is usually not the result of one single, well-identifiable factor. Instead, it is the \textit{interaction} between many factors that cause a situation to get out of control, while each factor alone does not necessarily constitute the source of disaster \textit{per se} \cite{Loveparade}.
\par
The following paragraphs offer some simple rules that can help one to improve the safety of mass events. These rules should not be considered as replacement of existing regulations, but can nevertheless complement more sophisticated procedures, and are easy to apply for all types of crowding, ranging from large mass events to daily commuting at train stations and airports.

\subsubsection{Design of the environment}
When mass events are planned, one simple rule to keep in mind is that counter-flows, merging flows, and crossing flows generate frictional effects and coordination problems among neighboring people, which create local peaks of density that increase the likelihood of congestion. Even though this is a well-established fact in academic circles, it remains a frequent source of crowd disasters~\cite{Loveparade}. Likewise, the topology of the environment plays a crucial role to ensure a smooth flow of people. Avoiding bottlenecks is probably the most important recommendation. The problems implied by bottlenecks have been demonstrated with many computer simulations, \textit{regardless} of the underlying model~\cite{Moussaid_2011, Hoogendoorn_2005, Helbing_2000}. Bottlenecks exacerbate the physical pressure among neighboring individuals, and create the initial perturbations that might trigger crowd turbulence. Furthermore, bottlenecks produce long queues and excessive waiting times. Therefore, while people naturally keep reasonable distances between each other, they tend to become impatient and reduce inter-individual distances when their perceived progress towards the goal is too slow. This causes a vicious circle: The increasing local density reduces the flow through the bottleneck, and the reduced flow further increases the local density. Importantly, bottlenecks do not only result from the static environment, but can also be created dynamically by other external factors. Ambulances, police cars, people sitting on the ground, lost belongings, or temporary fences often create bottleneck situations that were not initially foreseen. Another less intuitive side-effect is that the temporary widening of pathways can be dangerous as well, because people typically make use of the additional space to overtake those who walk in front of them. This creates bottleneck situations when the pathway narrows again to its normal width. Finally, recent computer simulations have revealed that sharp turns increase the physical pressure around the inner edge of the bend~\cite{Moussaid_2011}. Even though turnings are not necessarily dangerous, they constitute potential zones of danger in case of unexpected overcrowding.

\subsubsection{Early detection of problems}
When crowd turbulence begins, the situation typically quickly gets out of control. Detecting this danger early enough is crucial, but remains a challenge. Clear signs of emergency, such as people falling or calling for help usually happen at a stage that is already critical and does not enable a timely response. The most accurate indicator of danger is the \textit{local density} level. Recent studies have revealed the existence of local density thresholds above which the danger to the crowd is significantly increased~\cite{Moussaid_2011}: Stop-and-go waves tend to emerge beyond 2-4 people/m$^2$, while crowd turbulence is likely to occur above 4-7 people/m$^2$, depending on the average body diameter. The problem is that local densities are hard to measure without a proper monitoring system, which is not available at many mass events. Nevertheless, crowd managers can make use of simple signs as a proxy for local density. For example, one may observe the frequency and strength of body contacts among neighboring people. Typically, when the frequency of involuntary body contacts increases - often perceived as pushy behavior - this indicates that pressure in the crowd builds up and pressure relief strategies are needed. Crowd managers should also be alerted by the emergence of stop-and-go-waves, a self-organized phenomenon characterized by alternating moving and stopping phases, which is visible to the naked eye~\cite{Helbing_2007, Moussaid_2011}. As this stop-and-go pattern is related with a reduced flow of people, the density may quickly increase due to the vicious circle effect described in the previous section, so that measures for pressure relief must be taken. Finally, behaviors such as people climbing walls, overcoming fences, or disrespecting conventional routes - often interpreted as relentless and aggressive behaviors - may indicate escape attempts. They should be considered as a serious warning signal of a possible, upcoming crowd disaster.

\subsubsection{Information flows and communication}
Another major element in minimizing the risk of accidents is to set up an efficient communication system. A reliable flow of information between the visitors of an event, the organizers, security staff, police and ambulance is crucial for coordination. In fact, visitors often become aware of developing congestion too late because of the short-range visibility in crowded areas. This may impair orientation, which further increases the problems in the critical zone. In such situations, calm but clear loud speaker announcements informing visitors and well visible, variable message signs can let people know why they have to wait and what they should do. A robust communication system often constitutes the last means to gain control before a situation gets out of hand. It is also important that crowd managers have a reliable overview of the situation. Existing technologies offer useful monitoring systems that allow for real-time measurements of density levels and predictions of future crowd movements (also see section~\ref{sec:apps} describing the use of smartphone applications for inferring crowd density). While such sophisticated monitoring systems are among the most efficient ways to keep track of the situation, most events are not yet using them. The minimum necessary requirement, however, is video monitoring covering the relevant zones of the event, from which the organizers can precisely evaluate the crowd movements and detect signs of upcoming congestion.

\subsection{Apps for saving lives}  
\label{sec:apps}

Pre-routing strategies are an important element of crowd management as well. For this, a well-functioning information and communication system is crucial. While classically, the situation in the crowd is observed by means of surveillance cameras, helicopters or drones, and local police forces, there is an array of crowd sensors available today, based on video, WiFi, GPS, and mobile tracking. Smartphones, in particular, offer a new way of collecting information about crowded areas and of providing advice to attendees of mass events.

\begin{figure}
\centering
\includegraphics[width=10cm]{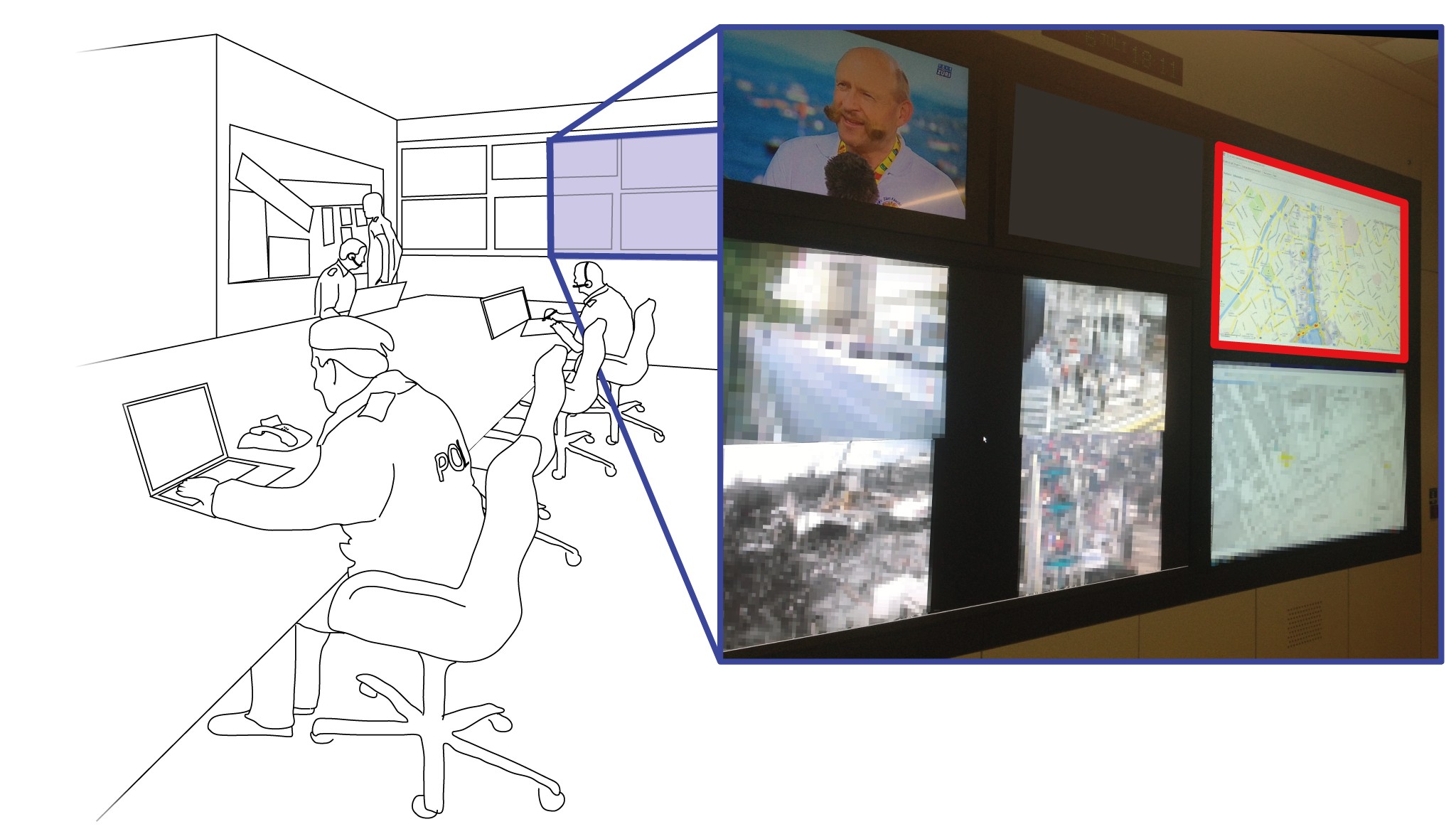}
\caption{Live view of crowd densities in the command and control center of the City Police Z{\"u}rich.}
\label{fig:c2c}
\end{figure}

\begin{figure}
\centering
\includegraphics[width=14cm]{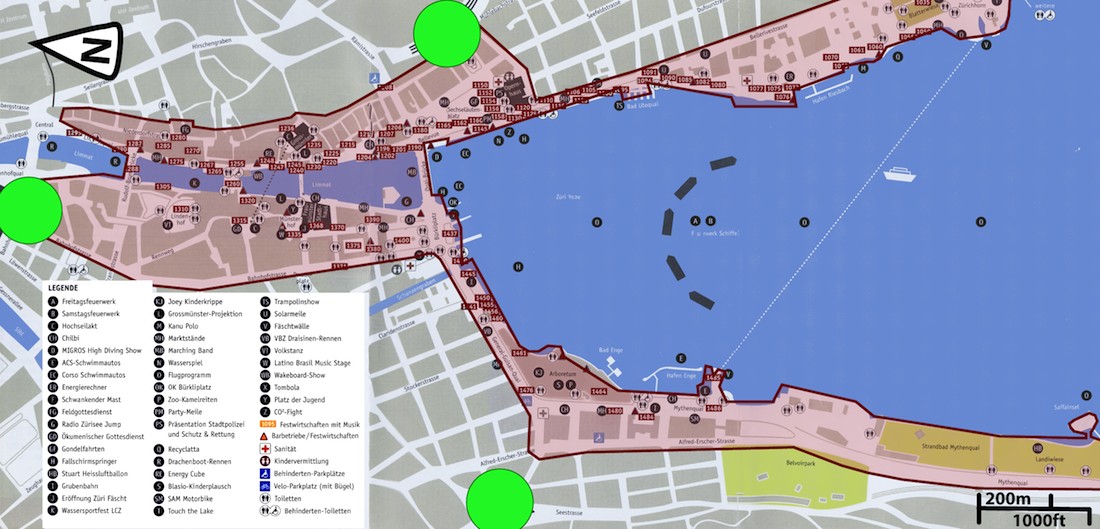}
\caption{Example of a festival site for the case of the ``Z{\"u}rif{\"a}scht'' in Z{\"u}rich, Switzerland, showing the perimeter (red) and the stage layout. The main train transportation hubs are indicated by green circles.}
\label{fig:zone}
\end{figure}

Advances in microelectronics have resulted in modern mobile phones that offer significant processing power, sensing capabilities, and communication features. In combination with a large market penetration they provide an excellent instrument for crowd monitoring. In particular, data can be acquired by deploying a mobile application that probes relevant modalities, e.g. the GPS or acceleration sensors. Adopting a participatory sensing scheme, macroscopic crowd parameters (e.g. density), mesoscopic parameters (e.g., collective behavior recognition, group detection, lane and queue detection, onset of panic prediction) and microscopic parameters of single individuals (e.g., modes of locomotion, velocity and direction, gestures) can be inferred.
\par
In safety critical and crowded environments such as major sports or religious events, instrumenting the environment with static sensors based on video or infrared cameras is currently the most reliable way to obtain a near 100\% sample of the crowd at a sub-second and sub-meter resolution required to accurately measure crowd pressure. However, smartphone apps can be used to infer a good approximation of crowd behavior at large scales, and can be used to pro-actively intervene before crowds reach critically high density levels.
\par
From a safety perspective, employing apps for large-scale events has at least three major advantages:
\begin{enumerate}
\item Broadband-communication allows to capture, transmit and centrally process data in near real-time and to extract and visualize relevant crowd parameters in a command and control center (Figure~\ref{fig:c2c}).
\item Offering bi-directional communication, safety personnel can send notifications, warnings, or even guide the user in case of an emergency situation. Incorporating the user's localization, geo-located messages increase the relevance for each user, helping to follow rules such as those outlined in section~\ref{sec:rulessafety}~\cite{franke2013participatory}.
\item Data, once collected, can be used to ``replay'' the event. The post hoc analysis is a critical step in the organization of an event and can reveal critical factors that should be addressed for future events. Although today a great deal of data is available for such analyses, event organisers and crowd managers are sometimes forced to manually scan through video material, reports from security services in the field, or feedback from individual visitors. Combining such heterogeneous sources is a tedious and error-prone task, thus making it difficult to reach a reliable assessment and situational awareness. Localizing users via smartphones potentially offers a mored accurate assessment and allows to capture crowd dynamics by aggregation.
\end{enumerate}
The use of smart phones for crowd monitoring is a very recent development. Different modalities such as bluetooth sensing~\cite{versichele2012use, stopczynskiparticipatory} or app-based GPS localization~\cite{wirz2013probing} have been used to capture collective dynamics such as flocking, crowd turbulence or mobility patterns patterns during large-scale events. In particular, individual location data can be aggregated using non-parametric probability density functions~\cite{wirz2013probing, wirz2012inferring} to estimate safety-relevant characteristics such as crowd movement velocity, density, turbulence or crowd pressure.
\begin{figure}
\centering
\includegraphics[width=14cm]{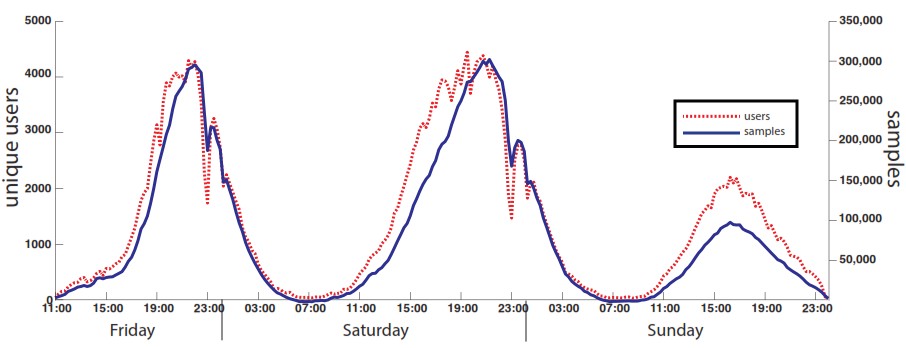}
\caption{Users and samples collected for 15 minute windows.}
\label{fig:usage}
\end{figure}

Here, we demonstrate the potential of location-data obtained from mobile phones during the Z\"uri F\"ascht festival 2013 in Switzerland (see Fig.~\ref{fig:zone}). The Z\"uri F\"ascht is a three-days event comprising an extensive program with concerts, dance parties, and shows. It is hosted in the city center of Z\"urich and is the biggest festival in Switzerland. Up to 2 million visitors have been estimated to attend the festival  over the course of three days. In 2013, 56,000 visitors downloaded the festival app, from which 28,000 gave informed consent to anonymously contributed their location data. Figure~\ref{fig:usage} shows the number of users simultaneously contributing their location data over the course of the event, and the amount of data samples collected. While collecting only a subsample from the entire crowd, previous work~\cite{wirz2013probing} showed a correlation coefficient greater than $0.8$ between the estimated density and actual density determined from video recordings.

\begin{figure}
\centering
\includegraphics[width=12cm]{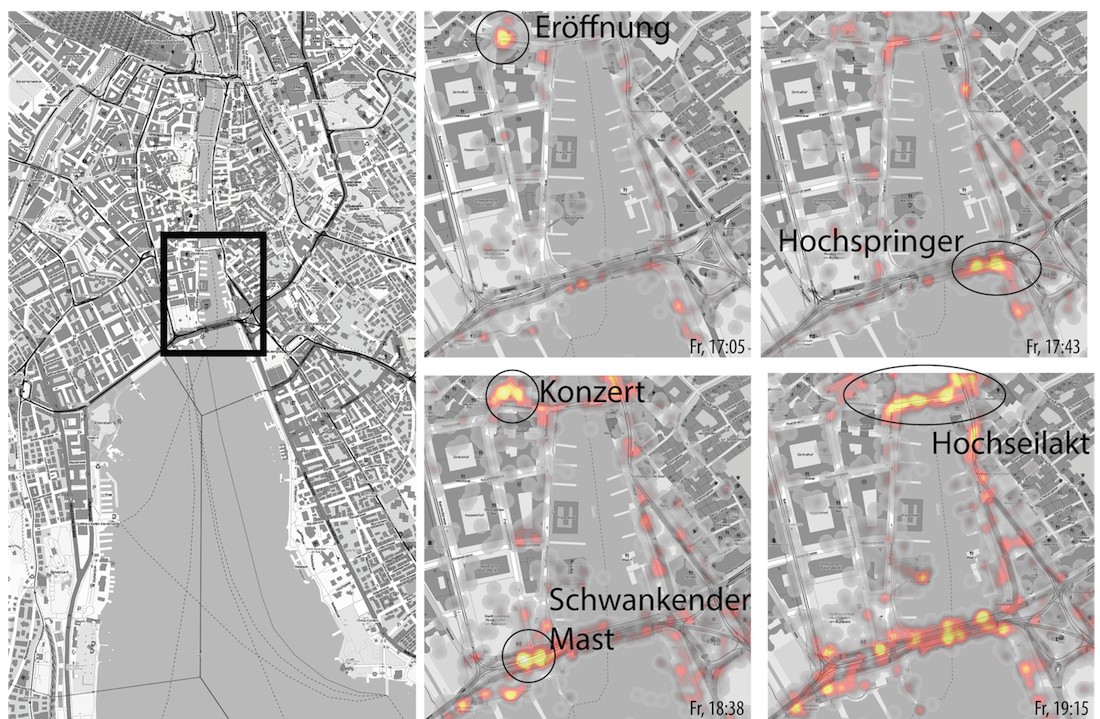}
\caption{Pedestrian densities during the Z{\"u}rif{\"a}scht festival for multiple attractions (the opening ceremony, a high diving competition, a concert, a high wire performance) at different times and locations.}
\label{fig:density}
\end{figure}

\begin{figure}
\centering
\includegraphics[width=12cm]{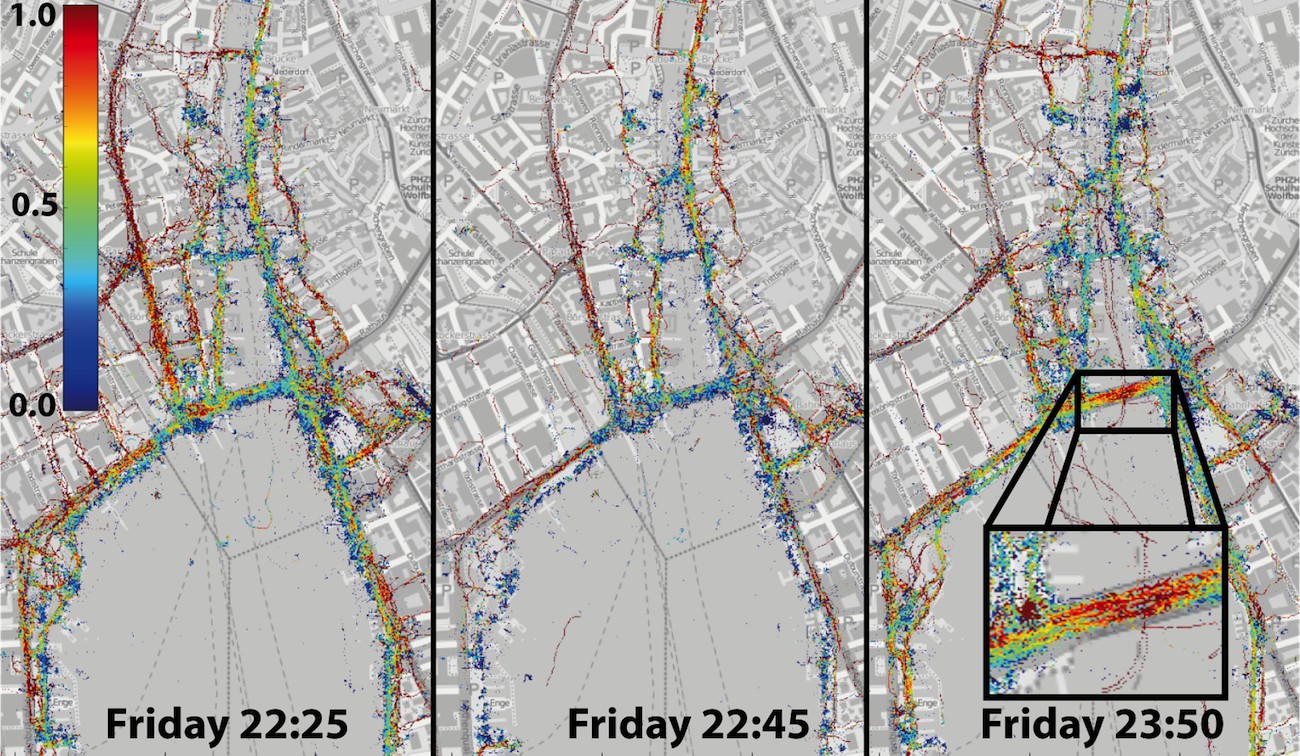}
\caption{Color-coded pedestrian speed before (left) during (middle) and after (right) the fireworks show.}
\label{fig:speed}
\end{figure}

\begin{figure}
\centering
\includegraphics[width=12cm]{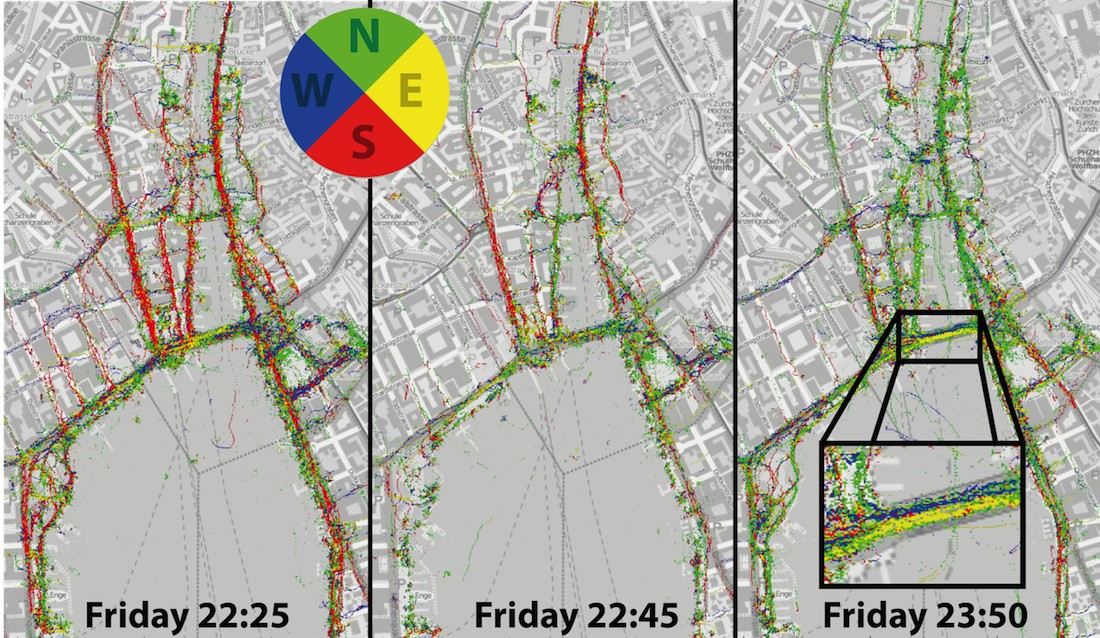}
\caption{Color-coded pedestrian direction before (left) during (middle) and after (right) the fireworks show.}
\label{fig:direction}
\end{figure}

The time-dependent density of visitors derived from the app data is shown in Figure~\ref{fig:density}. The velocity of festival visitors is represented in Figure~\ref{fig:speed} and the movement direction in Figure~\ref{fig:direction}. Together the three figures reflect the collective crowd dynamics. While Figure~\ref{fig:density} shows spectator densities for multiple events at different times, Figure~\ref{fig:speed} and Figure~\ref{fig:direction} show the mobility of the crowd before, during, and after the fireworks show at lake Z\"urich. Mobility patterns can be clearly identified. For instance, a strong northbound flow of the crowd can be observed before the event. Interestingly, lanes appear with a right-handed pedestrian traffic (highlighted box), which improves mobility. To support such kind of self-regulation, street signs (e.g., walking direction arrows) have been employed at other places, for instance at obstacles dividing the pathway.
\par
Several reality mining projects have demonstrated that it is not only possible to sense collective behaviors with mobile devices, but also to offer real-time feedback to the users via a return channel of the app. This allows crowd control staff to actively moderate the changes in the collective behavior. For example, the safety personnel may advise visitors to avoid crowded areas and suggest alternative routes. Such measures thus help improve the safety of crowds considerably.

\section{Crime, Terrorism, and War}
The quantitative study of crime and of violent conflict---both within and between nations---has a long tradition in the social sciences. With the growing availability of detailed empirical data in recent years, these topics have increasingly attracted the attention of the complexity science community. The strive to statistically characterize such data is not new---Richardson's seminal work on the size of inter-state wars, for example, dates back to 1948~\cite{Richardson_1948}. Recently, however, large datasets on both crime and conflict, often providing detailed information on each incident, its location and timing, have increasingly become available.\footnote{Detailed crime statistics world-wide are now often routinely released by the regional or national authorities that collect them. Large scale conflict data sets are usually collected and maintained by universities or research centers.} Such data has allowed the kind of statistical, system-level analysis we provide here to shed new light on a number of mechanisms underlying global crime and conflict patterns. The combined focus on crime and conflict in this section is intentional as both may be seen as breakdowns of social order, albeit of very different scale and arguably with very different origins.

\subsection{Fighting crime cycles: The cyclic re-emergence of crime and how to prevent it}
Crowd disasters are rare, but may kill hundreds of people at a time. In contrast to this, crimes are committed many times each day but usually involve only a few people in every incident. However, the impact of crime adds up and is considerable. In the US, for example, over 12,700 people were murdered in 2012 \cite{UCR_2012}. Economically, cybercrime, for example, results in hundreds of billions of dollars of damage world-wide every year.
\par
\begin{figure}
\centering
\includegraphics[width=0.87\columnwidth,clip]{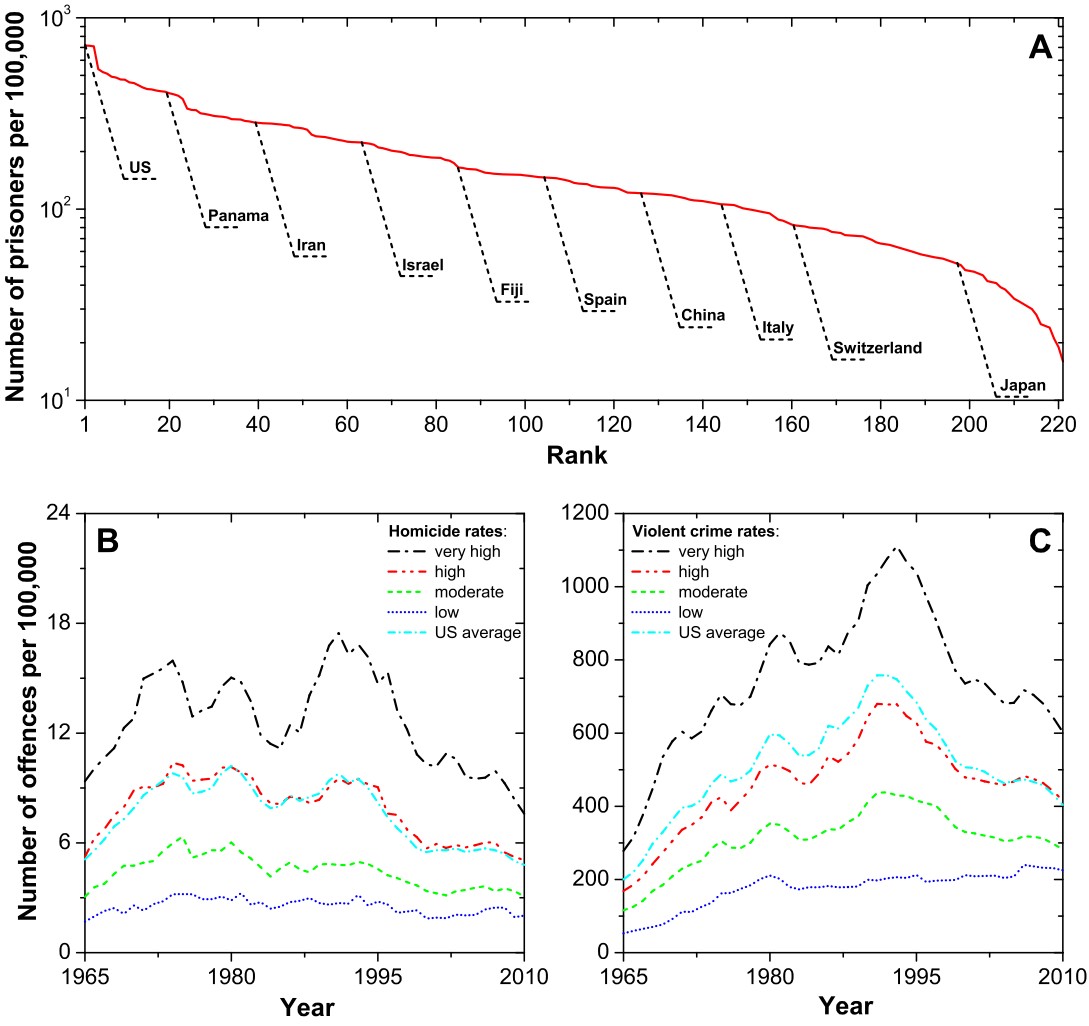}
\caption{(A) Empirical data on the imprisoned population world-wide (the full list of countries if available upon request). (B) number of homicides and (C) violent crimes in the US per 100,000 population. In addition to the average for the US (cyan), panels (B) and (C) show averages of US States with very high (black), high (red), moderate (green) and low (blue) crime rates. The four categories are obtained through grouping US States by their average crime rate over the period 1965-2010 and then assigning the 25\% most affected States to the first category, the 25\% next most affected States to the second category, etc. Examples of US States with very high crime rates are Washington D.C. and New York. States with low crime rates are, for instance, North Dakota and Utah.}
\label{data}
\end{figure}

To contain criminal activities, societies resort to various mechanisms punishing behavior that deviates from generally accepted norms in intolerable and destructive ways. The percentage of the population imprisoned in a country is an indication for the severity of state punishment (see Fig.~\ref{data}A). Remarkably, the US has the highest incarceration rate in the world~\cite{ICPS_2013}.\footnote{Since the beginning of the privatization of the the US prison services in the 1970s the number of prisoners in the US has increased 8-fold from about 300,000 to 2.4 million (US Bureau of Justice Statistics, http://www.bjs.gov/index.cfm?ty=pbse\&sid=5, accessed April 16, 2014). In international comparison this implies that with only a twentieth of the world population, the US accounts for a quarter of the world wide prison population~\cite{Economist_2013}.} Nevertheless, its crime rates remain very high by international comparison and are much higher than in many other industrialized countries~\cite{UNODC_2011}.\footnote{It is also interesting to note that a single prisoner annually costs more than the pay of a postdoctoral researcher. For a recent study on US prison costs per inmate see~\cite{VERA_2012}.} In particular, in connection with the ``war on drugs'', apparently more than 45 million arrests were made, yet it was recently declared a failure. In fact, US Attorney General Eric Holder concluded: ``Too many Americans go to too many prisons for far too long, and for no truly good law enforcement reason.''~\cite{Guardian_2013}.

But why do crime deterrence measures fail? According to standard rational choice theories of crime, they should not. The economic model of crime assumes that criminals are utility maximizers who optimize their payoffs $\pi$ under restrictions and risk~\cite{becker_jpe68}. There are gains $g>0$ (either material or physical or both) to be made with little effort, but there is also the threat of being sanctioned with a punishment fine $f>0$. If $s$ is the probability for an individual to commit a crime, and $p$ is the probability for the criminal act to be detected, the payoff is
\begin{equation}
\pi=s(g-pf) \, .
\label{rational}
\end{equation}
It is clear from Eq.(~\ref{rational}) that, at a given detection rate $p$, higher punishment (larger $f$) should reduce crime. However, this simple argument clearly neglects strategic interaction and numerous socio-economic factors, all of which play a crucial role in the evolution of criminal activity.
\par
In reality, increasing levels of deterrence do not necessarily reduce the level of crime. Instead, one often finds crime cycles. The homicide rate in the United States, for example, began to rise steadily in the 1960s and oscillated at around 8 to 10 homicides per 100,000 for 20 years. It only started to significantly decline in the early 1990s (see Fig.~\ref{data}B). In fact, such fluctuations in the frequency of crime are visible across a wide and diverse spectrum of criminal offenses, suggesting that cyclic recurrence is a fundamental characteristic of crime (see Fig.~\ref{data}C). This observation is notable, in particular, as different criminal offenses---homicide and theft, for example---are thought to arise from different kinds of motivations: homicides are usually assumed to be related to manifestations of aggression, whereas theft is most often economically motivated~\cite{Friedmann_1988}.
\par
Criminological research has identified a number of factors that influence the recurrence of crime, its trends and variation across regions and countries. Beside structural factors, such as unemployment~\cite{Raphael_2001,Lin_2008} and economic deprivation~\cite{LaFree_1999}, a number of studies have highlighted the influence of demography~\cite{Zimring_2007}, youth culture~\cite{Curtis_1998,Johnson_2006}, social institutions~\cite{LaFree_1998} and urban development~\cite{Barker_2012}. Others have pointed to the influence of political legitimacy~\cite{LaFree_1999}, law enforcement strategies~\cite{Corsaro_2009, McGarrell_2010}, and the criminal justice system~\cite{Zimring_2007}. Some recent studies in criminology, in fact, argue that trends in the levels of crime may be best understood as arising from a complex interplay of such factors~\cite{Barker_2012, Gomez_2006}. Very recent empirical work has further shown that social networks of criminals have a distinct impact on the occurrence of crime~\cite{Papachristos_2012, Papachristos_2013}, thus highlighting the complex interdependence of crime and its social context.
\par
Despite these advances the situation for policy makers to date remains rather opaque, since fully satisfactory explanations for the recurrence of crime are still lacking. Few studies fully recognize the complex interaction of crime and its sanctioning~\cite{short_pnas10, short_pre10, d2013criminal, perc2013und}. Analyzing this relationship, however, is of critical importance when trying to explain why, contrary to expectations~\cite{becker_jpe68}, increased punishment fines do not necessarily reduce crime rates~\cite{doob_cj03}. To understand its social context, crime should be considered as a social dilemma situation: it would be favorable for all if nobody committed a crime, but there are individual incentives to do so. As a consequence crime may spread, thereby creating ``tragedies of the commons''~\cite{hardin_g_s68}. Large-scale corruption and tax evasion in some troubled countries or mafia and drug wars are examples of this.
\par
If nobody is watching, criminal behavior (such as stealing) may seem rational to the individual, since it promises a considerable reward for little effort. To suppress crime it seems plausible to alter the decision situation formalized in Eq.(~\ref{decision}) such that the probability $p$ of catching a criminal times the imposed fine $f$ eliminates the reward $g$ of the crime, i.e.
\begin{equation}
g - pf < 0 \, .
\label{decision}
\end{equation}
According to this, whether it is favorable or not to commit a crime critically depends both on the fine $f$ and on the probability of catching a criminal $p$. Note that in particular the non-trivial interdependence of crime and the resources invested into detecting and punishing it should strongly affect the resulting spatio-temporal dynamics.
\par
In the simple, evolutionary game-theoretical model we discuss next,\footnote{The game discussed here was originally presented in Ref.~\cite{perc2013und}. Please refer to the original publication for further details.} the probability of detecting criminal activities is explicitly related to the inspection effort invested by the authorities. The game is played between ``criminals'' ($s_x = C$), punishing ``inspectors'' ($s_x = P$), and ``ordinary individuals'' ($s_x =O$) neither committing crimes nor participating in inspection activities. Each player tends to imitate the strategy of the best performing neighbor.\footnote{The imitation probability for two players $x$ and $y$ is given as $q=(1+\mathrm{exp}[(P_y-P_x)/K])^{-1}$, where $P_x$ and $P_y$ are the payoffs of the players and $K$ sets the intensity of selection.} The game is staged on a $L\times L$ square lattice with periodic boundary conditions, where individuals play the game with their $k=4$ nearest neighbors.  Criminals, when facing ordinary individuals, make the gain $g\ge 0$. When facing inspectors, however, criminals obtain the payoff $g-f$, where $f\ge 0$ is the punishment fine. If faced with each other, none of two interacting criminals obtains a benefit. Ordinary people receive no payoffs when encountering inspectors or other ordinary individuals. Only when faced with criminals, they suffer the consequences of crime in form of a negative payoff $-g \le0$.  Inspectors, on the other hand, always bear the cost of inspection, $c \ge 0$, but when catching a criminal, an inspector receives a reward $r\ge 0$, i.e. the related payoff is $r-c$. The introduction of dimensionless parameters is possible in the form of ``(relative) inspection costs'' $\alpha=c/f$, the ``(relative) temptation'' $\beta=g/f$, and the ``(relative) inspection incentive'' $\gamma=r/f$, which leads to the following payoffs:
\begin{eqnarray}
\pi_O &=& - \beta N_C , \\
\pi_P &=& \gamma N_C -\alpha, \\
\pi_C &=& \beta N_O + (\beta - 1) N_P ,
\end{eqnarray}
Herein, $N_O$, $N_P$, and $N_C$ are the numbers of ordinary individuals, inspectors and criminals among the $k=4$ nearest neighbors. Monte Carlo simulations are performed as described in the Methods section of  Ref.~\cite{perc2013und}.

\begin{figure}
\centering
\includegraphics[width=0.9\columnwidth,clip]{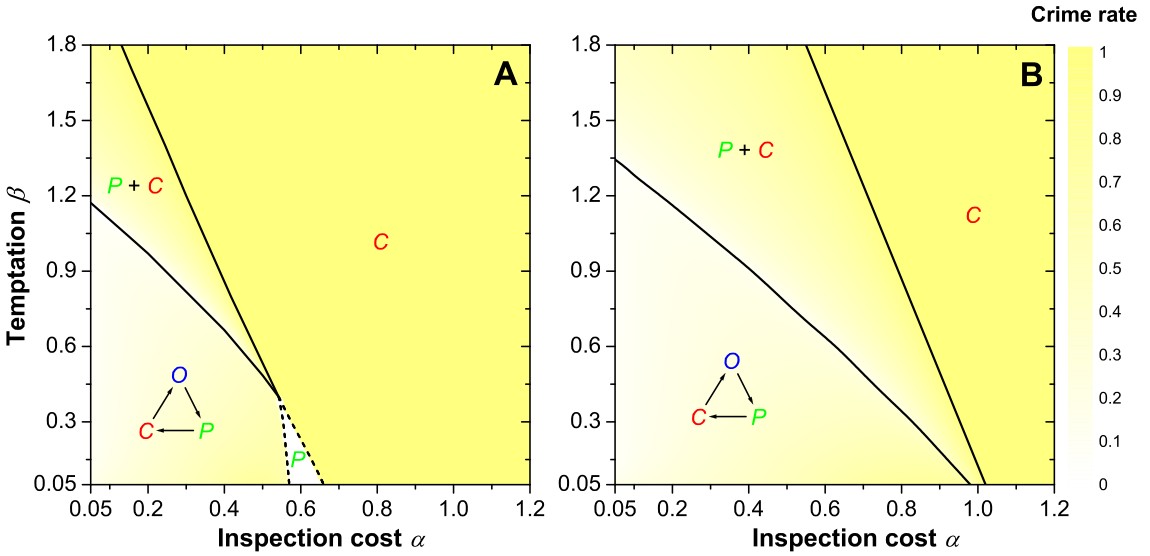}
\caption{Phase diagrams, demonstrating the spontaneous emergence and stability of the recurrent nature of crime and other possible outcomes of the competition of criminals ($C$), ordinary individuals ($O$) and punishing inspectors ($P$). The diagrams show the strategies remaining on the square lattice after sufficiently long relaxation times as a function of the (relative) inspection costs $\alpha$ and the (relative) temptation $\beta$, (A) for inspection incentive $\gamma=0.5$ and (B) for $\gamma=1.0$. The overlayed color intensity represents the crime rate, i.e. the average density of criminals in the system. (Figure adapted from Ref. \cite{perc2013und}).}
\label{phase}
\end{figure}

The collective spatio-temporal dynamics of this simple spatial inspection game is complex and counter-intuitive, capturing the essence of the crime-fighting problem well. One can observe various phase transitions between different kinds of collective outcomes, as demonstrated in Fig.~\ref{phase}. The phase transitions are either continuous (solid lines in Fig.~\ref{phase}) or discontinuous (dashed lines in Fig.~\ref{phase}), and this differs markedly from what would be expected according to the rational choice equation~(\ref{decision}) or the well-mixed model~\cite{tsebelis_rs90}, namely due to the significant effects of inspection and spatial interactions. Four different situations can be distinguished:
\begin{enumerate}
\item dominance of criminals for high temptation $\beta$ and high inspection costs $\alpha$,
\item coexistence of criminals and punishing inspectors ($P+C$ phase) for large values of temptation $\beta$ and moderate inspection costs $\alpha$,
\item dominance of police for moderate inspection costs $\alpha$ and low values of temptation $\beta$, but only if the inspection incentives $\gamma$ are moderate [see panel (A)], and
\item cyclical dominance for small inspection costs $\alpha$ and small temptation $\beta$, where criminals outperform ordinary individuals, while these outperform punishing inspectors, and those win against the criminals ($C+O+P$ phase) (supplementary videos of different routes towards the emergence of the cyclic phase are available at \textcolor{blue}{youtube.com/watch?v=pH9l-2h6PRo} and \textcolor{blue}{youtube.com/watch?v=gVnCN3a9ki8}).
\end{enumerate}
\par
\begin{figure}
\centering
\includegraphics[width=0.85\columnwidth,clip]{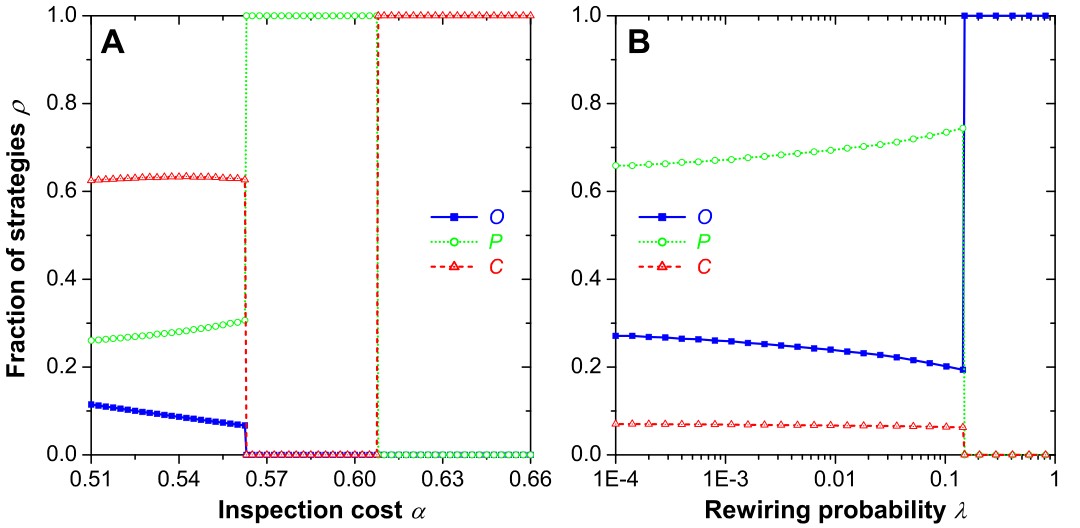}
\caption{A representative cross-section of the phase diagram presented in Fig.~\ref{phase}A (A), and the robustness of crime cycles against the variation of network topology (B). These results invalidate straightforward gain-loss principles (``linear thinking'') as proper description of the relationship between crime, inspection, and punishment rates. (A) For $\beta=0.2$ and $\gamma=0.5$, increasing $\alpha$ initially leaves the stationary densities of strategies almost unaffected, while subsequently two discontinuous first-order phase transition occur. (B) The social interaction networks were constructed by rewiring links of a square lattice of size $400 \times 400$ with probability $\lambda$. As $\lambda$ increases, due to the increasing interconnectedness of the players, the amplitude of oscillations become comparable to the system size. A supplementary video is available at \textcolor{blue}{youtube.com/watch?v=oGNOmLognOY}. (Figure adapted from Ref.~\cite{perc2013und}).}
\label{cross}
\end{figure}

Figure~\ref{cross} provides a more detailed quantitative analysis, displaying in panel (A) the sudden first-order phase transitions from the $C+O+P$ phase to the pure $P$ phase, and from this pure $P$ phase to the pure $C$ phase, as inspection costs $\alpha$ increase. In panel (B), we also observe a sudden, discontinuous phase transition, emerging as neighboring individuals are rewired. This introduces a small-world effect. We observe that conflicts escalate in amplitude until an absorbing phase is reached. It follows that the exact impact of each specific parameter variation depends strongly on the location within the phase diagram, i.e. the exact parameter combination. General statements such as ``increasing the fine reduces criminal activity'' tend to be wrong, which contradicts the widely established point of view. This may explain why empirical evidence is not in agreement with common theoretical expectations, and it also means we need to reconsider the traditional perspective on crime.

The model reviewed here, therefore, demonstrates that that ``linear thinking'' is inadequate as a means to devise successful crime prevention policies. The level of complexity governing criminal activities in competition with sanctioning efforts appears to be much greater than often assumed so far. Our results also reveal that crime is likely to be recurrent when there is a gain associated with criminal activity. While our model is, of course, stylized, it can nonetheless help to shed light on the counterintuitive impact of punishment on the occurrence and, in particular, recurrence of crime.
\par
Our model also highlights that crime may not primarily be the result of activities of individual criminals. It should be rather viewed as result of social interactions of people with different behaviors and the collective dynamics resulting from imitative interactions~\cite{Papachristos_2012, Papachristos_2013}. In other words, the emergence of crime may not be well understood by just assuming a ``criminal nature'' of particular individuals---this picture probably applies just to a small fraction of people committing crimes. In fact, criminal behavior in many cases is opportunity-driven or arises as a result of social interactions that get out of hand. This suggests that changing social context and conditions may make a significant contribution to the reduction of crime, which in turn has relevant implications for policies and law.
\par
It is commonly assumed that we have come a long way in deterring crime. Many different types of crime are punished all over the world based on widely accepted moral norms, and there exist institutions, organizations, and individuals who do their utmost to enforce these norms. However, specific deterrence strategies may sometimes have no, or even unintended, adverse effects. Recent research finds, for example, that even though video surveillance usually allows for a faster identification of criminal delinquents, it rarely leads to a significant decrease in criminal activities in the surveyed area~\cite{Gill_2005}. In fact, it does not always lead to an increased individual perception of safety~\cite{Zehnder_2011}, even though this is a key arguments for its widespread use.
\par
Other recent findings suggest that certain police strategies might even be outright counter-productive. Based on the assumption that removing the leader of a criminal organization will disrupt its social network, police often attempt to identify and arrest him or her. A recent study analyzing cannabis producer networks in the Netherlands, however, shows that this strategy may be fundamentally flawed: all network disruption strategies analyzed in the study did not disrupt the network at all or, worse, increased the efficiency of the network through efficient network recovery~\cite{Sloot_2014}.

These examples clearly highlight that an insufficient understanding of the complex dynamical interactions underlying criminal activities may cause strong adverse effects of well-intended deterrence strategies. A new way of thinking, maybe even a new kind of science for deterring crime is thus sorely needed---in particular one that takes into account not just the obvious and similarly linear relations between various factors, but one that also looks particularly at the interdependence and interactions of each individual and its social environment. One then finds that this gives rise to strongly counter-intuitive results that can only be understood as the outcome of emergent, collective dynamics. This is why complexity science can make important and substantial contributions to the understanding and containment of crime.

\subsection{Dynamics of terrorism and insurgent conflict}
\begin{figure}[t]
\centering
\includegraphics[width=14.0cm]{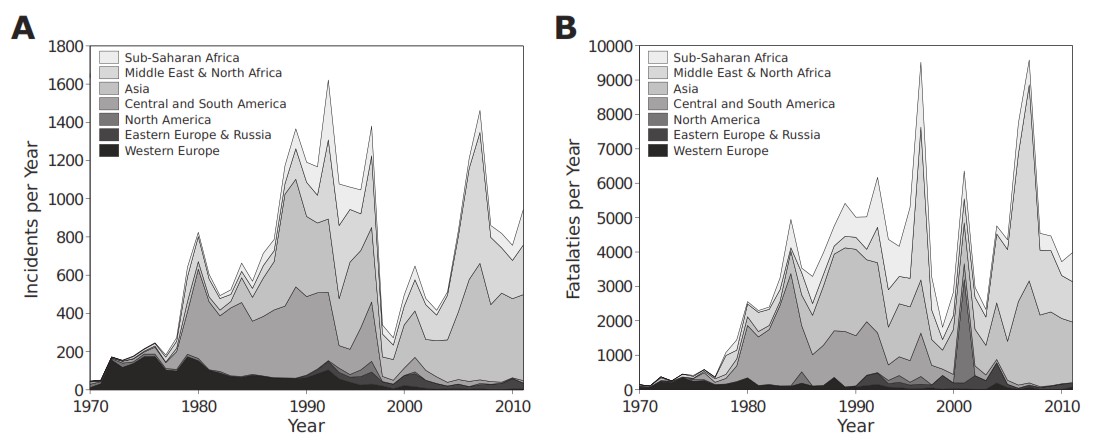}
\caption{International terrorism disaggregated by region. (A) number of events and (B) fatalities per year. The data were obtained from the Global Terrorism Database (GTD) (http://www.start.umd.edu/gtd/) and represent incidents clearly identified as terrorist attacks that lead to at least one fatality.}
\label{conflict:terrorism}
\end{figure}

The past decade has seen a resurgence of international terrorism and several violent civil conflicts. The insurgencies that followed the US-lead interventions in Afghanistan and Iraq have caused civilian casualties in the order of hundreds of thousands while the US-declared ``war on terror'' has not significantly curbed international terrorist activity (Fig.\ref{conflict:terrorism}). In fact, in the political arena, critics argue that international interventions may themselves serve as triggers for terrorist attacks while others suggest the exact opposite, i.e. that only the decisive interventions and aggressive measures against terrorist organizations have stopped a much more severe escalation of international terrorism.

There are no studies to date that can convincingly support either of the above views and establish a clear causal relationship between international terrorist activities and the ``war on terror''. At the level of individual conflicts, however, there are a number of studies that have helped to clarify the link between interventions and levels of violence~\cite{Lyall_2009, Braithwaite_2012, Linke_2012, Lyall_2013}. These studies in particular emphasize reactive or ``tit-for-tat'' dynamics as a fundamental endogenous mechanism responsible for both escalation and de-escalation of conflict. In the context of Iraq, one can find such reactive dynamics for both insurgent and coalition forces~\cite{Braithwaite_2012, Linke_2012}. Other research has identified similar dynamics in the Israeli-Palestinian conflict where both the Israeli and Palestinian side were found to significantly react to violent attacks of the opposing side~\cite{Haushofer_2010, Clauset_2010}. Establishing the (causal) effect of actions by one conflict party on the reactions of the other faction(s) thus appears central to a systemic understanding of endogenous conflict processes.
\par
\begin{figure}
\centering
\includegraphics[width=12.0cm]{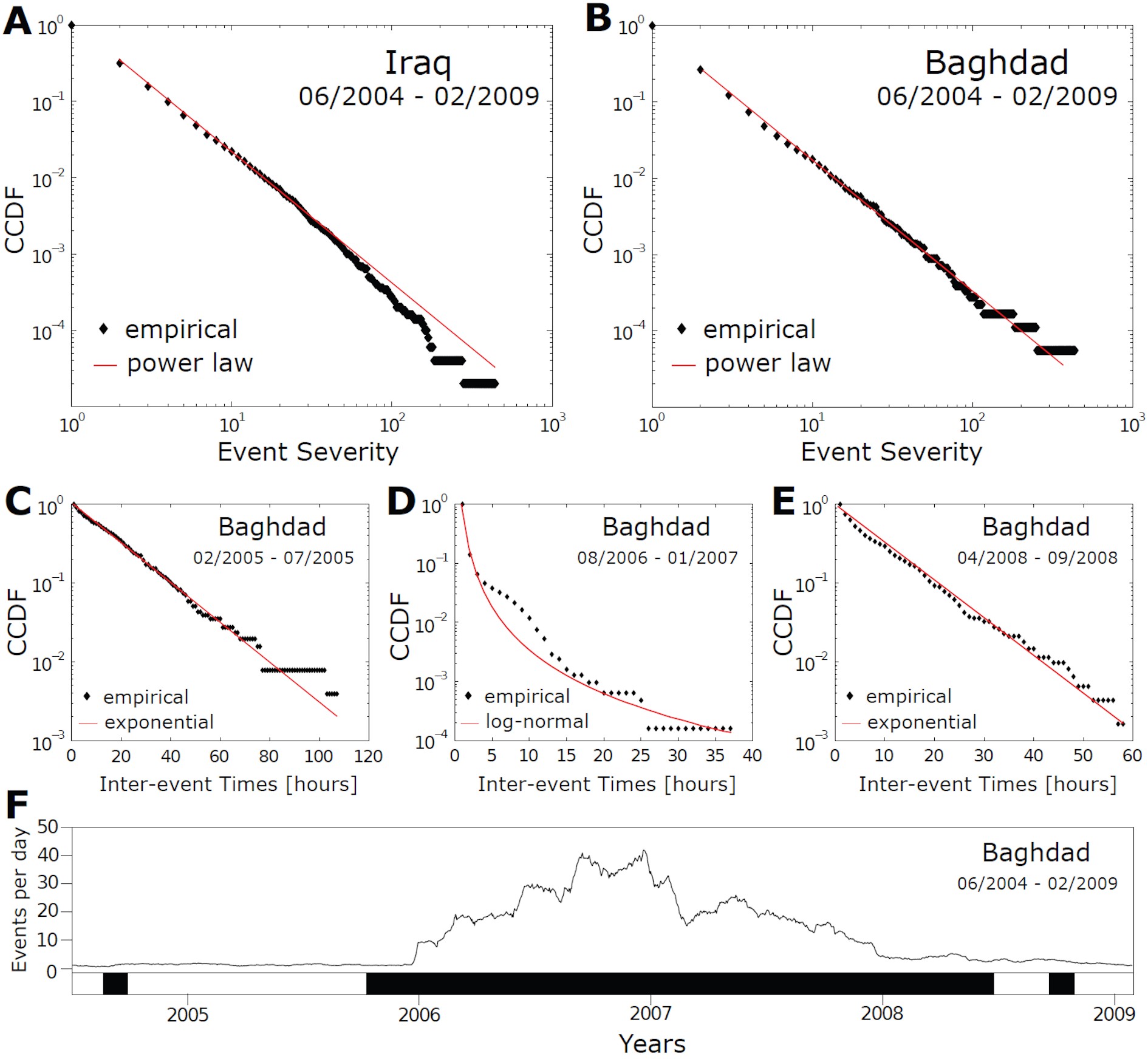}
\caption{Empirical severity and timing signatures of conflict events during the war in Iraq (2004--2009). The complementary cumulative distribution functions (CCDFs) of the event severity in Iraq (A) and Baghdad (B) follow robust power laws ($\alpha=2.48\pm0.05$, $x_0=2$, $p<0.001$ and $\alpha=2.57\pm0.05$, $x_0=2$, $p<0.001$ respectively). Panels (C) through (E) show the CCDF of inter event times in a 6 months time window for different periods of the conflict in Baghdad. In 2006 ad 2007, the timing of events correlates significantly in time, i.e. the signature of inter-event times (D) significantly deviates from the exponential timing signature that is characteristic for uncorrelated event timing. In contrast, prior to 2006 (C) and after 2007 (E) event timing is indistinguishable from that of a random process with exponentially distributed inter-event times. (F) Shows the number of events per day in Baghdad for the whole period considered. The black bar indicates where an Anderson-Darling test rejects the hypothesis of exponential distribution of inter-event times at a 5\% significance level for moving windows of 6 months, i.e. where the event timing signature is non-trivial. Note that timing analysis was performed for the greater Baghdad area to avoid spurious timing signatures of unrelated events (see also the discussion in the text). The data were obtained from the Guardian website~\cite{Rogers_2010}; fit lines shown were estimated using maximum likelihood estimation~\cite{Clauset_2009}.}
\label{conflict:Iraq}
\end{figure}

There is complementary statistical evidence pointing to relatively simple relationships underlying aggregate conflict patterns. Empirical data of insurgent attacks, for example, exhibits heavy tailed severity distributions and bursty temporal dynamics~\cite{Clauset_2007, Bohorquez_2009}. The complementary cumulative distribution function (CCDF) of event severities is usually found to follow a power law (see Fig.~\ref{conflict:Iraq}):
\begin{equation}
\mathrm{Pr}(X\ge x) \sim x^{\alpha}, \quad x\ge x_0
\end{equation}
where $x_0$ is the lower bound for the power law behavior with exponent $\alpha$. Note that such statistical regularities can, for example, be used to estimate the future probability of large terrorist events~\cite{Clauset_2013}. Bursty temporal dynamics may be characterized by considering the distribution of inter-event times, i.e the length of the time intervals between subsequent events. In a structureless dataset, that is in a dataset where the timing of events is statistically independent, the CCDF of inter-event times simply follows an exponential distribution. The deviation of timing signatures from an exponential shape---for example log-normal or stretch exponential signatures---thus points to significant correlations between the timing of subsequent events.\footnote{Non-stationary time series exhibit non-exponential timing signatures even in the absence of correlations because the CCDF of a mixture of exponential is already heavy-tailed. In any statistical analysis this non-stationarity thus has to be either explicitly modeled (for instance using parametric methods) or the analysis has to be restricted to sufficiently small time windows in which the dynamics can be assumed to be (approximately) stationary.} Researchers have offered competing explanations for these system-level statistical regularities. The power laws in the severity of attacks, for example, have been linked to competition of insurgent groups and security forces for control of the state~\cite{Clauset_2007}, but also to a positive feedback loop between the size and experience of terrorist organizations or insurgent groups and the frequency with which they commit attacks~\cite{Clauset_2012}.

It is important to note that the choice of spatial and temporal units of analysis in such aggregate, system-level analysis may strongly affect results. Choosing empirically motivated and sufficiently small spatial units of analysis avoids problems arising from considering time series of potentially unrelated or only very weakly related incidents. In the context of Iraq, for example, the violence dynamics in the Kurdish dominated North are generally quite different from those in Baghdad or the Sunni triangle. The same applies to the selection of suitable time periods, in which the same conflict mechanisms produce aggregate patterns. In fact, the non-parametric analysis of event timing signatures rests on the assumption that the conflict dynamics in the period analyzed are (approximately) stationary~\cite{Bohorquez_2009}. Timing analyses are therefore usually performed for shorter time intervals---for example 6 months---where the overall intensity of conflict does not significantly change (see Fig.~\ref{conflict:Iraq}).
\par
While such statistical signatures point to similarities in conflict patterns across cases and help identify general conflict mechanisms, detailed case studies on the micro-dynamics of violence suggest that mechanisms at the group- or individual-level may be very conflict- and context-specific~\cite{Donnay_2014}. A number of micro-level mechanisms that all contribute to the aggregate violence patterns are likely to be present in many conflicts (see, for example Ref.~\cite{Weidmann_2013}). The effect of (local) contact between warring factions on levels of violence, for instance, has been of central interest in conflict studies. Examining violence in Jerusalem between 2001 and 2009, a recent study addresses this question using an innovative combination of formal computational modeling and rich, spatially disaggregated data on violent incidents and contextual variables~\cite{Bhavnani_2013}. Past research has given empirical support for two competing perspectives: The first assumes that intermixed group settlement patterns reduce violence, as more frequent interactions enable rivals to overcome their prejudices towards each other and thus become more tolerant. The other argues that group segregation more effectively reduces violence, given less frequent contact and fewer possibilities for violent encounters (see Ref.~\cite{Bhavnani_2013} for a detailed overview of the literature supporting the two perspectives).
\par
The study on Jerusalem now demonstrates that, in fact, both perspectives can be reconciled, if one acknowledges that the social or cultural distance $\tau$ between groups effectively mitigates the effect of spatial proximity. Spatial proximity, in other words is necessary but not sufficient to trigger local outbreaks of violence between members of different groups---only in a situation where tensions between groups are high a chance encounter on the street would likely trigger a violent outbreak. Formally, the study assumes that the probability of any chance encounter to result in violence depends on an individual's violence threshold $\Gamma$, which changes as a function of personal exposure to violence (decreases) and the threat of police intervention (increases). Whether this disposition to engage in violence translates into a violent incident then is conditional on the social or cultural distance $\tau$, such that the higher $\tau$, the more likely it is that violence ensues:
\begin{equation}
\mathrm{p}(x)=\left(1+\mathrm{exp}\left [\frac{-(\tau-\Gamma)}{\lambda}\right ]\right)^{-1}
\end{equation}
Here $\lambda$ sets the scale of the probability of violence to erupt, the smaller $\lambda$ the more abrupt the transition from no violence to violence as a function of $\tau-\Gamma$.
\par
One can quantitatively show that this local, individual-based mechanism can indeed explain a substantial part of the aggregate violence dynamics in Jerusalem~\cite{Bhavnani_2013}. Moreover, considering counter-factual scenarios the model allows to analyze how different levels of segregation corresponding to proposed ``futures'' for Jerusalem would affect violence levels in the city. Conceptually, the study highlights the complex inter-dependence of micro-level conflict processes. In fact, not only the effect of (local) contact is conditional on socio-cultural distance: conflict (or its absence) in turn also affects the socio-cultural distances between conflict parties.
\par
One key difficulty in research on the micro-dynamics of conflict is focusing on the ``correct'' theoretically and empirically plausible mechanisms that drive violence. In the study on Jerusalem, for example, one specific mechanism was shown to account for a significant part of the spatial patterns of violence in the city. This and similar studies are powerful in the sense that they explicitly test a particular causal micro-macro-link. Similarly, one may reverse this process and try to infer micro-processes based on the macro-patterns. Recently, such causal inference designs have been extended to be applicable to micro-level conflict event data. A novel inferential technique called \emph{Matched Wake Analysis} (MWA)~\cite{Schutte_2014} estimates the causal effect of different types of conflict events on the subsequent conflict trajectory. Using sliding spatial and temporal windows around intervention events guarantees that the choice of unit of analysis does not systematically bias inferences, a problem known as the modifiable areal unit problem (MAUP)~\cite{Kulldorff_1997, Openshaw_1979}. For robust causal inference the method uses statistical matching~\cite{Iacus_2012} on previous conflict trends and geographic covariates. In order to the estimate average treatment effects for one type of event (\textit{treatment}) relative to another one (\textit{control}) the method uses a Difference-in-Differences regression design. Formally, it estimates how the number of dependent events after interventions, $\mathrm{n}_{\mathrm{post}}$, changes compared to the number of dependent events prior, $\mathrm{n}_{\mathrm{pre}}$, as a function of interventions:\footnote{\textit{treatment} is a Boolean marking whether the intervention is of the \textit{treatment} or \textit{control} type.}
\begin{equation}
\mathrm{n}_{\mathrm{post}}=\beta_0+\beta_{1}\cdot\mathrm{n}_{\mathrm{pre}}+\beta_{2}\cdot treatment +\mathrm{u}
\end{equation}
In this expression $\beta_{2}$ is the estimated average treatment effect of the treated, our quantity of interest (see Ref.~\cite{Schutte_2014} for further details on the method).
\par
Such inferential methodology is especially useful when testing specific causal hypotheses, in particular those that may otherwise be difficult to detect against the background of large-scale violence unrelated to the hypothesized mechanism. We have previously emphasized the importance of reactive dynamics for an understanding of endogenous conflict processes. One such example is the effects of counterinsurgency measures on the level of insurgent violence. The direction of these effects is generally disputed, both theoretically and empirically. One line of reasoning suggests that hurting civilians indiscriminately will generally cause reactive support for the military adversary~\cite{Kocher_2011, Linke_2012}. Another line of thinking suggests that the deterrent effect of such measures leads to less support for the adversary~\cite{vanCrefeld_2008, Lyall_2009}.
\par
\begin{figure}
\centering
\includegraphics[width=14.0cm]{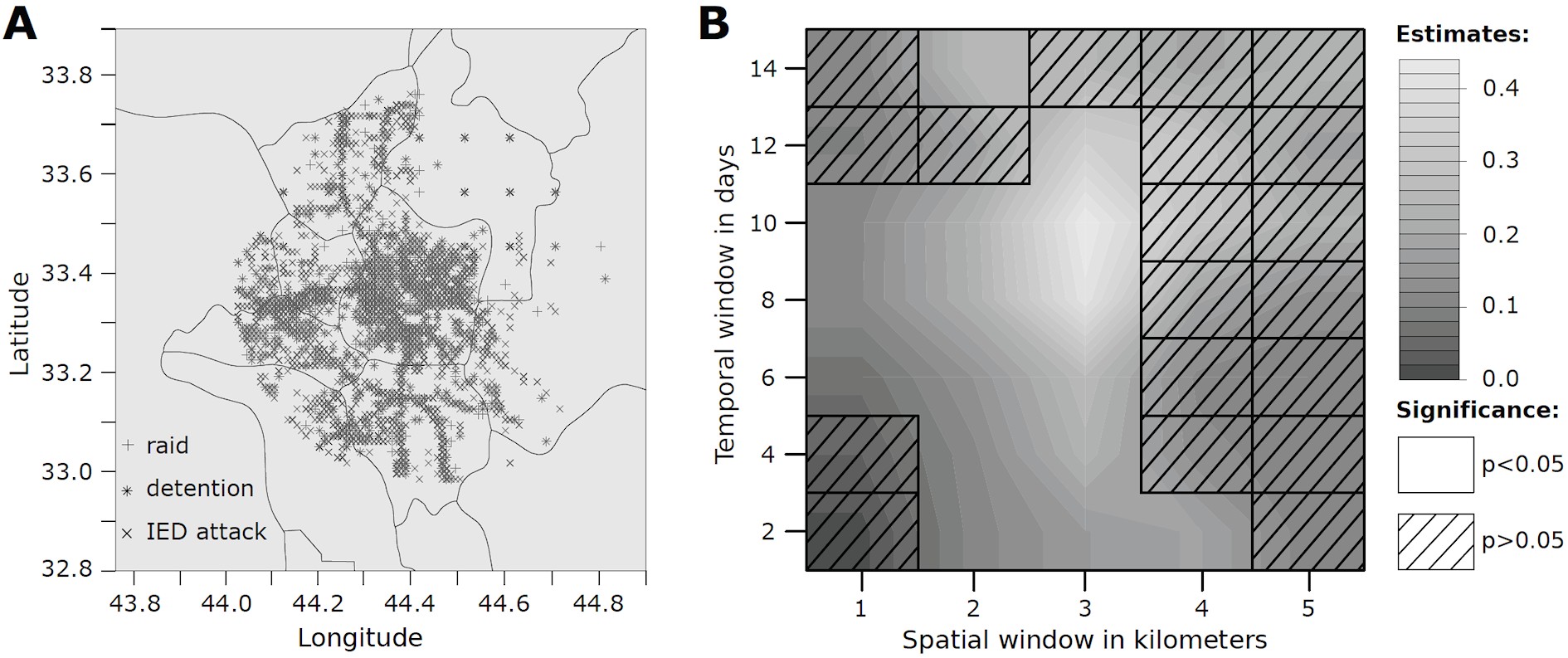}
\caption{Reactive violence dynamics in Iraq. (A) Map of the greater Baghdad area for the period 01/2004 to 03/2006 showing the sample of conflict events used in our analysis. (B) Estimated change in the number of attacks with Improvised Explosive Devices (IEDs) following raids in comparison to those following detentions. After matching on spatial covariates and trends in IED attacks preceding the interventions, the causal effect of raids vs. detentions was estimated with a Difference-in-Differences regression design~\cite{Schutte_2014}. Note the noticeable positive effect (lighter colors) in the interpretable (non-shaded) areas of the plot at distances up to about 3 kilometers and 10-12 days: on average for every 2-3 raids one more IED attack is observed compared to less heavy-handed interventions. Data used for this analysis are significant action (SIGACT) military data (see for example Ref.~\cite{Linke_2012}).}
\label{conflict:matchedwake}
\end{figure}

We use the MWA technique here to directly test this effect empirically by focusing on the initial insurgency (beginning of 2004 to early 2006) in Iraq. Specifically, we focus on the greater Baghdad area---the focal area of violence---and consider three frequent types of events. \emph{Raids} refer to surprise attacks on homes and compounds, in which the military suspects insurgents or arms caches. The relative heavy-handedness and the high probability of disturbing or harming innocent bystanders sets these operations apart from more selective \emph{detentions}. In turn, to attack incumbent forces insurgents frequently rely on \emph{Improvised Explosive Devices} (IEDs) and civilian support in manufacturing, transporting, and planting them. If civilians were really more inclined to support the strategic adversary after having been targeted in raids, one would expect more IED attacks to take place after and in the spatial vicinity of raids compared to detentions or arrests under otherwise comparable conditions. The opposite should be true if deterrence (by raids) was the mechanism at work. As visible in Fig.~\ref{conflict:matchedwake}, raids lead to higher levels of subsequent IED attacks compared to detentions: our estimates suggest that 2-3 raids, lead to one additional IED attack in the direct spatial vicinity within up to 10-12 days. This result thus lends empirical support to the notion that civilians support the adversary in reaction to indiscriminate violence. Based on such improved methodology for studying micro-level processes in conflicts, it is possible to gain a better and systematic understanding of the driving forces behind violence.

The research discussed in this section highlights that the contributions of complexity science and empirical, case-oriented conflict research to the study of conflict are in many ways complimentary. A systemic perspective can serve to identify abstract general relationships that help shed light on aggregate empirical patterns---much like statistical physics, which uses the self-averaging properties of large $N$ systems to study system level properties that emerge from complex microscopic interactions~\cite{Clauset_2012}. The in-depth and theory-driven analysis of selected cases in the literature on micro-dynamics of conflict on the other hands, tries to systematically identify these detailed micro-level mechanisms. While there has recently been a noticeable shift towards the study of micro-dynamics of conflict with significant conceptual and technical progress~\cite{Donnay_2014}, it is important to note that historically much of the social science literature on conflict has mainly analyzed aggregate country- or system-level data.
\par
The fundamental challenge to date remains bridging the divide between these two perspectives. In theory, statistical analysis at the level of aggregate distributions can offer unique systematic insights for the whole ensemble with detailed  micro-level inference revealing the multitude of mechanisms underlying the ensemble properties. Yet in practice, the two research strands often exist rather disconnected and insights from one do not enter into the analysis or models of the other. While there may in principle be a systematic limit to the generalizability of insights from the macro to the micro level and vice versa, the current situation is clearly a consequence of lack of engagement with the respective ``other'' literature.
\par
We are convinced that in light of the increasing availability of detailed conflict event datasets and the emergence of Big Data on conflicts, complexity science can make a very substantial contribution to conflict research. Its impact and relevance will be more substantial the more we tap into the unique insights on both the micro- and macro-dynamics of conflict the social science literature already has to offer.

\subsection{Interstate wars and how to predict them}
If conflicts are difficult to stop, then it should be of utmost importance to prevent their occurrence in the first place. However, any hope of preventing them rests on the ability to forecast their onset with a certain level of accuracy.
\par Unfortunately, forecasting conflicts has remained largely elusive to scholars of international relations. Historical studies of single wars lack out-of-sample predictive power~\cite{pevehouse1999serbian, schrodt1996using, bosler2011oracle}. More systematic approaches focusing on the \emph{conditions} that tend to lead to war (e.g., arms races~\cite{glaser2000causes}, long-standing territorial rivalries~\cite{huth1998standing}, or large and rapid shifts in power~\cite{powell2004inefficient, chadefaux2011}) rely on indicators that are coarse in time (typically yearly), and therefore tend to miss the crucial steps of the escalation of tensions and the timing of the conflict outbreak~\cite{beck2004theory, de2004untangling, BecKinZen00, gleditsch2011forecasting}. Efforts at finer-grained codification of geopolitical tensions are labor-intensive and costly~\cite{azar1980conflict, mcclelland1984world} and limited in time \cite{weidmann2010predicting,IIspecialIssue2012}. More systematic coding mechanisms using computer algorithms~\cite{king2003automated, GDELT} also have a limited time span, as do prediction markets~\cite{wolfers2006prediction, berg2008prediction}. In addition, event-based data tend to miss the subtleties of international interactions: on the one hand, the absence of an event may be as important as its occurrence. Moreover, seemingly large events need not be cause for alarm whereas small events may greatly matter. The main obstacle to testing our ability to forecast conflicts, in other words, has been the lack of measures of tensions that are both fine-grained in time and cover a large time-span~\cite{holsti1963value, newcombe1974development, choucri1974forecasting}.
\par To fill this gap, geopolitical tensions were estimated by analyzing a large dataset of historical newspaper articles~\cite{chadefaux2014}.  Google News Archive---the world's largest newspaper database with over 60 million pages\footnote{The database spans more than 200 years and includes a uniquely comprehensive set of historical newspapers.}---was used to search the text of every article for mentions of a given country together with a set of keywords typically associated with tensions (e.g., crisis, conflict, antagonism, clash, contention, discord, fight, attack, combat).\footnote{E.g.: `Sir Edward Grey seeking conference to avert a general \emph{conflict}. \emph{France} and Italy agree' [The New York Times, 08-28-1914]. See details in Ref.~\cite{chadefaux2014}.}. A mention of a country together with any of the pre-specified keywords in a given week resulted in an increased estimated level of tensions for that country in that week. This procedure was repeated for every week from January 1$^{\textrm{st}}$ 1902 to December 31$^{\textrm{st}}$ 2011, and for every country included in the Correlates of War dataset~\cite{cowStates}.

\par  The resulting dataset is a fine-grained and direct proxy for the evolution of tensions in each country, with more than one hundred years of weekly time series for 167 countries. Moreover, by relying on journalists' \emph{perceptions} of international tensions, some of the pitfalls associated with event-data can be avoided, since contemporaries will process events in view of their respective context and likely consequences, instead of simply classifying them in preexisting categories.

\par
\begin{figure}[t]
\centering
 \includegraphics[width=0.4\textwidth]{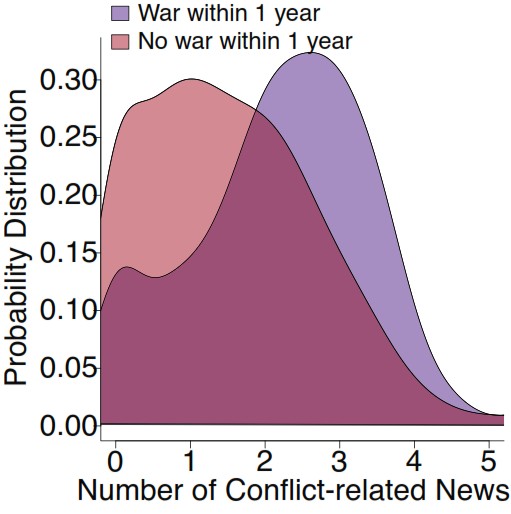}
 \caption{Estimated probability distribution of the (logged) number of conflict-related news for all countries and all weeks since 1902. The curve ``War within 1 year'' refers to the distribution of conflict-related news for those countries involved in an interstate war within the next twelve months (i.e., a conflict with at least 1,000 battle deaths involving two or more states).}
 \label{Fig:PlotChadefaux_density}
\end{figure}

\begin{figure}[t]
\centering
 \includegraphics[width=0.4\textwidth]{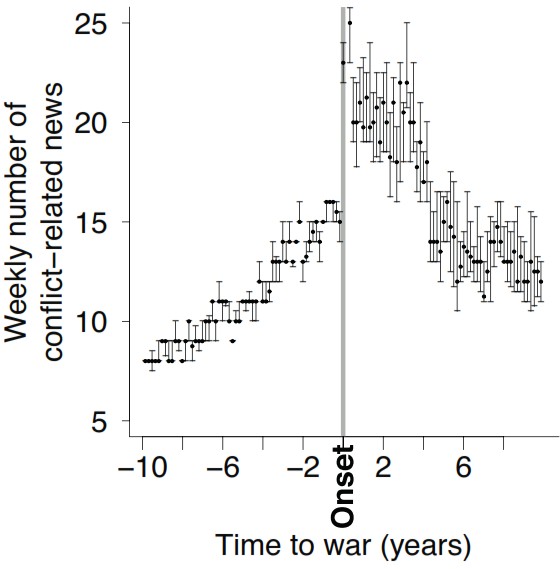}
 \caption{Observed number of conflict-related news items prior and after the onset of interstate wars. Each vertical line is a boxplot of the bootstrapped median number of conflict-related news prior to all interstate wars since 1902. The number of news tends to rise well ahead of the onset of conflict, and to remain relatively high thereafter, reflecting growing concerns prior to a conflict and lingering ones after its onset~\cite{chadefaux2014}.}
 \label{Fig:Chadefaux_news_time}
\end{figure}

 Reports about tensions in the news were found to be significantly higher prior to wars than otherwise (Fig.~\ref{Fig:PlotChadefaux_density}), which implies that news reports convey valuable information about the likely onset of a conflict in the future~\cite{chadefaux2014}.
In fact, reports about geopolitical tensions typically increase well ahead of the onset of conflict (Fig.~\ref{Fig:Chadefaux_news_time}), therefore potentially giving decision-makers ample early warnings to devise and implement a response to help prevent the outbreak.

\par Even when controlling for a multitude of explanatory variables that have been identified as relevant in the literature (e.g., regime type, relative power, military expenditure), the sheer number of conflict-related news provides earlier warning signals for the onset of conflict than existing models. This hypothesis was tested more formally by using only information available at the time (out-of-sample forecasting). Using such data, the onset of a war within the next few months could be predicted with up to 85\% confidence. Keyword-based predictions significantly improved upon existing methods. These predictions also worked well before the onset of war---more than one year prior to interstate wars---giving policy-makers significant additional warning time~\cite{chadefaux2014}.
\par  Here we report on an additional finding suggesting the importance of uncertainty prior to conflict as an early warning signal for war. The outbreak of conflict is rarely unavoidable. While countries often prepare for potential conflicts well in advance, the escalation of tensions typically reflects a failed bargaining process, and news articles mirror this escalation by reporting on these developments. Yet, escalation is rarely linear, and its outcome remains uncertain until the very end. A negotiated solution can always be reached before hostilities start. As a result, proximity to the conflict will probably increase not only observers' (journalists) worries about the outcome of the process, but may also generate diverging opinions and scenarios about the likely outcome of the process. As a result, we should observe not only an increase over time in the number of conflict-related news, but also an increase in the variance associated with these news.
\begin{figure}[t]
\centering
 \includegraphics[width=0.4\textwidth]{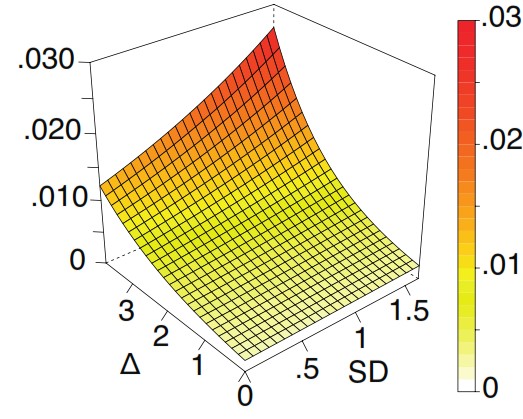}
 \caption{Estimated probability of the onset of interstate war within the next three months. $\Delta$news denotes the weekly change in the number of conflict-related news  (i.e., $\Delta$news$_{i,t}\equiv$ news$_{i,t}/(1+$news$_{i,t-1}$ for country $i$ and week $t$), and SD measures the moving standard deviation of the number of news (divided by the total number of news in the world to address non-stationarity problems), applying a one-year averaging moving window.}
 \label{Fig:chadefaux_InternationalWars}
\end{figure}

\par Indeed, we find that the probability of conflict significantly increases as a function of both the weekly change in the number of conflict-related news items ($\Delta$news) and their standard deviation within a window of one year (Fig.~\ref{Fig:chadefaux_InternationalWars}). This results goes beyond~\cite{chadefaux2014} and suggests not only that the total number of conflict-related news and its change over time can be used as early warning signals for war, but also that time-variability in the number of news could significantly improve upon existing forecasting efforts. This finding is also in line with~\cite{Scheffer2009}, since increased levels of variability can be used as early warning signals for critical transitions in a large number of physical, biological and socio-economic systems characterized by complexity and non-linearity (see also~\cite{HelbingEPJB2009}).

\section{Spreading of diseases and how to respond}
\begin{figure}
\centering
\includegraphics[width=1\columnwidth]{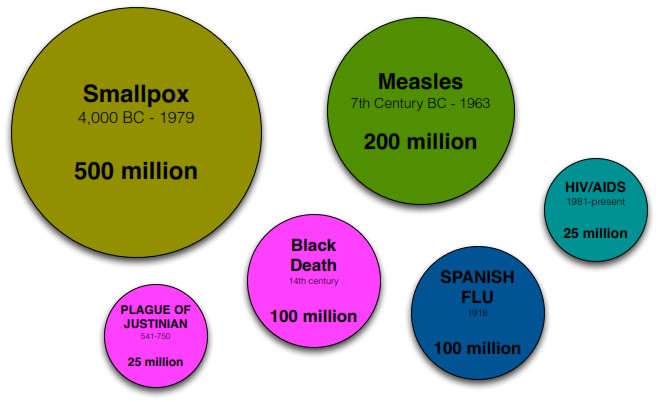}
\caption{Pandemics of the past and present. Global disease proliferation has been prevalent throughout human history. Before worldwide eradication in 1979, smallpox had claimed the largest death toll since its emergence approx. 5000 yrs. ago followed by measles which is believed to have killed a substantial fraction of native Americans in the times after Columbus. Both, the Justinian Plague and the Black Death were caused by bacterial diseases and each one killed more than half the European population in a few years. The 20th century Spanish flu claimed more victims that world war I and HIV/AIDS is the most recent globally prevalent non-curable infectious disease.}
\label{fig:Killers}
\end{figure}

In the previous sections, we have seen how crime and conflict can spread in space and time. We will now address another serious threat, the emergence and worldwide spread of infectious diseases. Pandemics have plagued mankind since the onset of civilization. The transition from nomadic hunter-gatherer life styles to settled cultures marked a point in history at which growing human populations, in combination with increasingly cultured life-stock, led to zoonotic infectious diseases crossing the intra-species barrier. New pathogens that could be sustained in human populations evolved, and eventually lead to infectious diseases specific to humans. It is thus no surprise that the greatest killer among viral diseases, the smallpox virus, emerged approx. 4000-7000 years ago~\cite{Hughes:2010fr}. It is estimated that smallpox alone has caused between 300 and 500 million deaths, and along with measles killed up to 90\% of the native American population in the years that followed contact with Europeans in the 15th century. Historic records indicate that various large scale pandemics with devastating consequences swept through Europe in the past two millennia. The Antonine Plague (smallpox or measles) killed up to 5 million people in ancient Rome and has been argued to be one of the factors that lead to the destabilization of the Roman Empire~\cite{bruun:2006}, while the Black Death (bubonic plague) swept across the European continent in the 14th century and erased more than 25\% of the European population~\cite{Hays:2005wu}. The largest known pandemic in history, the Spanish flu of 1918-19, caused an estimated 20-50 million deaths worldwide within a year, more than the casualties of World War I over the previous 4 years. This pandemic was caused by an influenza A virus subtype H1N1, a strain similar to the one that triggered the 2009 influenza pandemic (``swine flu''). Although all of these major events are unique and different from each other in many ways, a number of factors are common to all of them. First, increasing interactions with animal populations in farming increases the likelihood that novel pathogens are introduced to human populations. Second, larger and denser human populations increase the opportunity for pathogens to evolve the ability to be sustained in these populations, adapt to the new human host, and trigger local outbreaks. Third, increasing mobility, for instance driven by trade and commerce, promotes the spatial distribution of emergent pathogens, and generates the conditions for full-blown pandemics to follow a local outbreak.
\par
In recent decades, the threat of pandemics has been substantially reduced by a more advanced medical system, the development of sophisticated antibiotics, antiviral drugs, and vaccination campaigns. The greatest success along these lines is the worldwide eradication of smallpox in 1979 by a decade-long global vaccination campaign. Today, worldwide health surveillance systems for timely outbreak identification are in place, containing emergent infectious disease, and combating endemic diseases such as polio, tuberculosis, and malaria. However, although advances along these lines are promising news, modern civilization also contributes to the emergence of new pandemics. First, intensive animal farming or industrial livestock production yields a higher rate of intra-species barrier crossing and thus the emergence of potentially virulent human infectious diseases. Second, worldwide population increases quickly, having recently crossed the 7 billion mark. More than 50\% of the Earth's population live in dense metropolitan mega cities, thus providing ideal conditions for an emergent pathogen to be sustained by human-to-human contacts. Third, the modern world is extremely connected by intense long range traffic. For example, more than three billion passengers travel among four thousand globally distributed airports every year.
\par
Containing global pandemics is certainly a major concern to ensure stable socio-economic conditions in the world and anything that can potentially reduce the socio-economic impact of these events is helpful. In the last decade mathematically parsimonious SIR (susceptible-infected-recovered) models~\cite{anderson1991infectious} have been extended successfully by including social interaction networks, spatial effects, as well as public and private transport, and by more and more fine-grained epidemiological models~\cite{Rvachev:1985tb, Hufnagel:2004kt, Eubank:2004p1115, Ferguson:2006p509, Balcan:2010wn, VandenBroeck:2011dj}. One of the most relevant and surprising lessons of such models is that, in order to minimize the number of casualties and costs of fighting diseases in industrialized countries, it might be economically plausible for them to share vaccine doses with developing countries, even for free~\cite{Colizza:2007p1066}. In this way, the number of infections can be more effectively contained, particularly in the critical, early stage of disease spreading.
\par
In the following, we will focus on two aspects. Firstly, a better understanding and prediction of the spatio-temporal spreading of diseases on a global scale, which permits the development of more effective response strategies. Second, we will address the issue of information strategies to encourage voluntary vaccination of citizens, which might be more effective than attempting to enforce compulsory vaccination through law. In this context, it is useful to remember that each percentage point reduction in the number of people falling ill potentially benefits the lives of tens of thousands of people.

\subsection{Modelling disease dynamics}
\label{sec:epidemics1}
The development of mathematical models in the context of infectious disease dynamics has a long history. In 1766 Daniel Bernoulli published a paper on the effectiveness of inoculation against smallpox infections~\cite{Bernoulli:2004wg}. His seminal work not only contained the first application of the theory of differential equations, but was also published before it was known that infectious diseases were caused by bacteria or viruses and were transmissible. The prevalent scientific opinion in Bernoulli's time was that infectious diseases were caused by an invisible poisonous vapor, known as the miasma. Evidence existed, however, that inoculation of healthy individuals with degraded smallpox material incurred immunity to smallpox in some cases. Bernoulli's theoretical work shed light on the effectiveness of this procedure, a topic vividly discussed by the scientific elite of Europe at that time. In the beginning of the 20th century Kermack and McKendrick laid the foundation of mathematical epidemiology in a series of publications~\cite{Kermack:1927p414,Kermack:1933p405} and introduced the SIR (susceptible - infected - recovered) model which is still used today in most state-of-the-art computational models for the dynamics of infectious diseases (see also Refs.~\cite{ANDERSON:1979we,MAY:1979p382}).
\par
These types of models were designed to describe the time course of epidemics in single populations in which every individual is assumed to interact with every other individual at the same rate. The spatial spread of epidemics was first modeled by parsimonious reaction diffusion type equations~\cite{Fisher:1937p5494} in which the local nonlinear dynamics are combined with ordinary diffusion in space. The combination of local, initially exponential growth typical of disease dynamics, with diffusion in space, generically yields regular wave fronts and constant spreading speeds. This type of approach was successfully applied in the context the spread of the Black Death in Europe in the 14th century~\cite{NOBLE:1974p774}. Despite their high level of abstraction, these models provide a solid intuition and understanding of spreading processes. Their mathematical simplicity allows one to compute how key properties (e.g. spreading speed, arrival times, and pattern geometry) depend on system parameters~\cite{Murray:2002uo}.

\begin{figure}
\centering
\includegraphics[width=1\columnwidth]{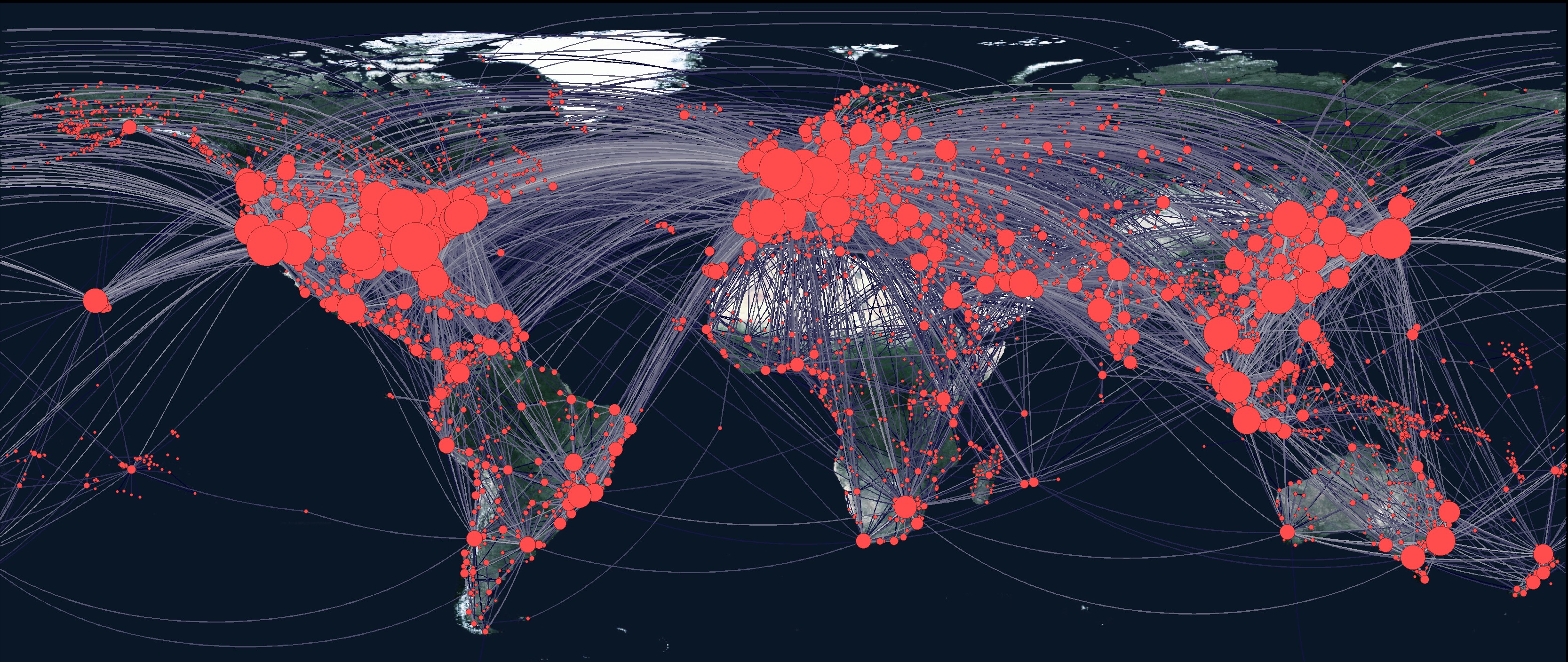}
\caption{Global connectivity in the 21st century: Heterogeneity and lack of scale. The network represents the worldwide air-transportation network connecting approx. four thousand airports worldwide. Symbol size quantifies airport capacity, a proxy for local population size. Two major characteristics of global connectivity are scale-free connectivity in terms of distance and strong heterogeneity. These factors yield spatio-temporal complexity of global disease dynamics.}
\label{fig:wan}
\end{figure}

One of the key problems in understanding global disease dynamics in the 21st century is that some of the fundamental assumptions in homogeneous reaction-diffusion models are not valid, not even approximately: (1) the host population is typically not distributed homogeneously in space and (2) global human mobility is characterized by a lack of scale which cannot be accounted for by local diffusion processes. This is illustrated in Fig.~\ref{fig:wan} which shows the worldwide air transportation network. This network connects approx. four thousand airports worldwide and transports more than three billion passengers every year. Recent examples have shown that the strong spatial heterogeneity of the human population in combination with the lack of scale in global mobility no longer lead to regular wave patterns of global disease spread. SARS (Severe Acute Respiratory Syndrome) in 2003 initially proliferated in China but only shortly after its arrival in Hong Kong cases were reported in Canada, the United States and Europe. In 2009, the H1N1 pandemic started in Mexico, quickly reached the US and subsequently arrived in Europe and elsewhere in the world. The spread of modern epidemics is spatially incoherent and chaotic, generically does note bear any metric regularity, and observed spatio-temporal patterns depend sensitively on the outbreak location. All these factors complicate the development of predictive tools that can forecast the time course of emergent infectious diseases.
\par
Despite these challenges, complexity science has produced major advances in modeling the dynamics of global epidemics. Increasing computer power and the availability of a new quality of pervasive data, in combination with important theoretical insights from network science, have enabled scientists to develop highly sophisticated computational platforms that can simulate large scale pandemics in silico. Modern super-computers allow running highly detailed agent-based simulations that model up to almost a billion host individuals, each one with a specific behavioral profile. State of the art computer models also employ insights that have recently been gained by analyzing pervasive data obtained in natural experiments on human interactions, in particular, about human mobility. Smart phones and geo-aware devices have produced data from which parameters can be extracted that previously had to be assumed or guessed in models for disease spread.
\par
The most successful computational frameworks for predicting disease dynamics on a global scale account for multiple factors which are believed to play a role in a specific context: demographic variation, mobility patterns that include the entire global air-traffic system as well as the short-scale, daily commuter movements in almost every country on the planet, detailed epidemiological data, and disease-specific mechanisms~\cite{VandenBroeck:2011dj, Balcan:2010wn, Ferguson:2006p509, Colizza:2007p1066, Hufnagel:2004kt, Rvachev:1985tb}.

\begin{figure}
\includegraphics[width=1\columnwidth]{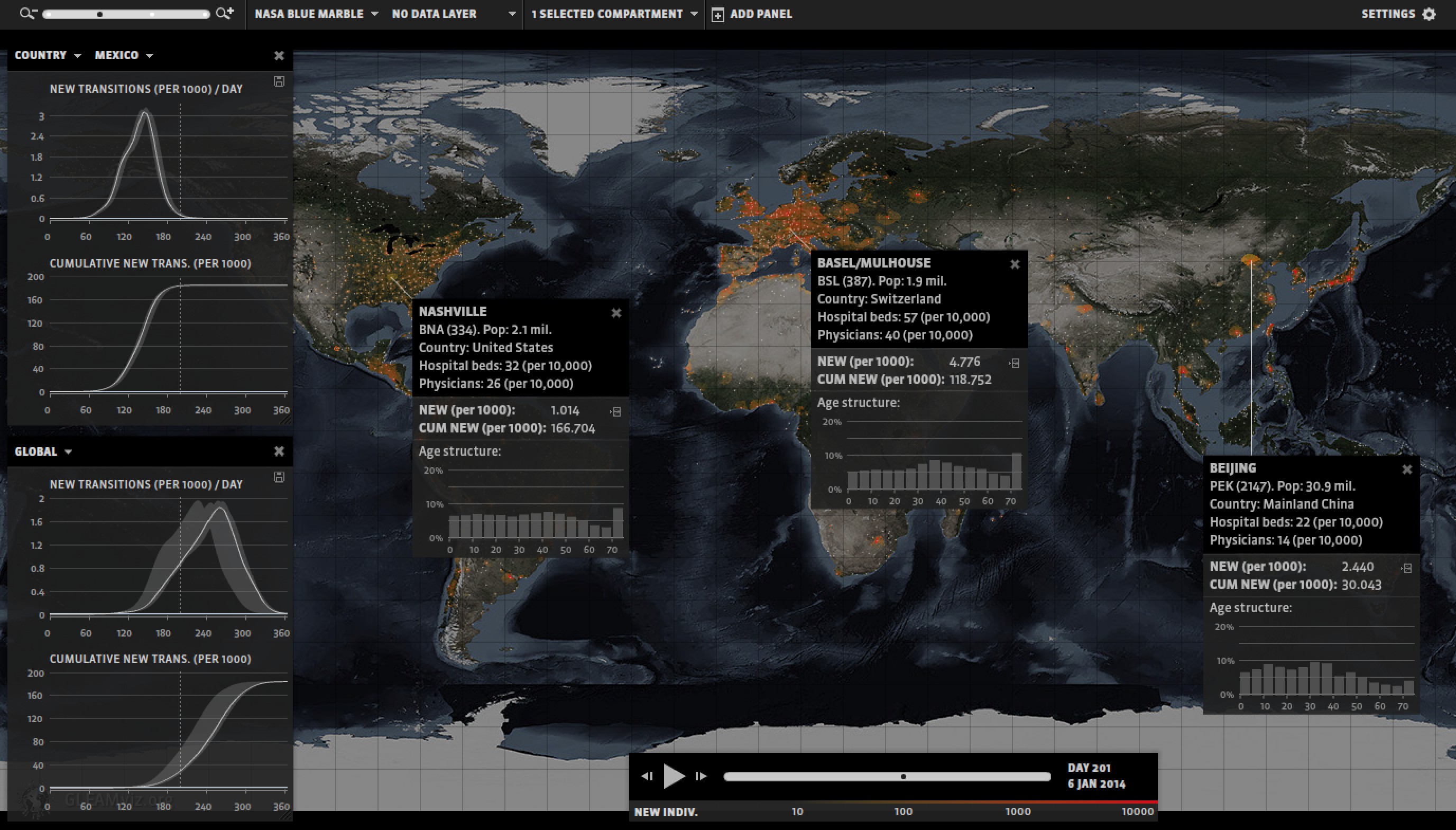}
\caption{GLEAM, the global epidemic and mobility simulation software. The image illustrates the user interface of the interactive simulation software (http://www.gleamviz.org/). Today, GLEAM represents one of the most sophisticated and powerful simulation frameworks for global disease dynamics and has been employed in a variety of contexts.}
\label{fig:gleam}
\end{figure}
A large scale operation in this direction is the Global Epidemic and Mobility (GLEAM) simulation framework~\cite{VandenBroeck:2011dj}. This tool has a client/server architecture that allows the simulation of global epidemics by accounting for the entire worldwide air-transportation network in combination with daily commuter traffic in most countries. The approach accounts for a diverse set of mobility patterns, e.g. bi-directional movements between origin and destination, or different mobility patterns for host individuals in different infectious states. Hhospitalization, vaccination, age structure at every location, and seasonality, can also be incorporated in the model setup (see Fig.~\ref{fig:gleam}. Because of the client/server architecture, a model can be defined on the client, and the simulation itself can be run on a remote super-computer in little time. A clear advantage of the GLEAM software is that entire sets of fully stochastic simulations can be performed in order to compute confidence intervals along with the most probable time course. In this respect the design of the GLEAM infrastructure is similar to modern weather forecast systems. The GLEAM project and related simulation approaches have become remarkably successful in reproducing observed patterns and predicting the temporal evolution of ongoing epidemics~\cite{Tizzoni:2012gx} for instance in predicting the time-course of the 2009 H1N1 pandemic.
\par
Interestingly, many of the predictive, computational models that exist reproduce similar dynamic features despite significant differences in their underlying assumptions and data~\cite{Ajelli:2010fg}. Furthermore, the level of detail these models incorporate also implies an abundance of parameters that either must be assumed, or in some cases guessed. Generically these parameters interact non-linearly in a given model, and despite their predictive power, it is difficult to assess which parameters are important and which are less significant. We currently lack a full understanding of the dynamic richness of these highly detailed models and therefore also of the phenomena they describe. Another drawback of purely detailed computational approaches is the fact that parameter values must be chosen and fixed to run a specific computer simulation. This is particularly challenging when new infectious agents emerge for which none of the vital disease specific parameters are known. Consequently, at the onset of an emergent epidemic even the most sophisticated simulation frameworks have to be employed with great caution because, despite their potential to model the phenomenon in a very detailed way, a significant level of uncertainty remains concerning the choices of sensitive parameters.
\par
It also remains unclear, how the multitude of factors shape the dynamics and how much detail is required to achieve a certain level of predictive fidelity. Most importantly, detailed computational models that incorporate all potentially relevant factors ab initio fail to reveal which factors are actually relevant, and which ones are not~\cite{May:2004el}. Gaining a better understanding of the phenomenon, however, is particularly important because based on a deeper understanding we can potentially predict certain aspects about the most likely time course of a global epidemic, even when a set of parameters is unknown. Along this line of reasoning, in~\cite{Brockmann_2013} the authors introduce the notion of effective distance to replace conventional geographic distance. This move simplifies the complexity of the spatio-temporal dynamics of global diseases to the point that even complex network-driven contagion phenomena can be understood in the framework of ordinary reaction-diffusion dynamics. In other words, when effective distance replaces geographic distance, spreading phenomena that appear to be complex in conventional geographic representations possess simple wave fronts and constant speeds. Thus, the key result of the study was that the observed complexity is predominantly caused by the inappropriate and antiquated use of purely spatial distance measures.
\par
\begin{figure}
\includegraphics[width=1\columnwidth]{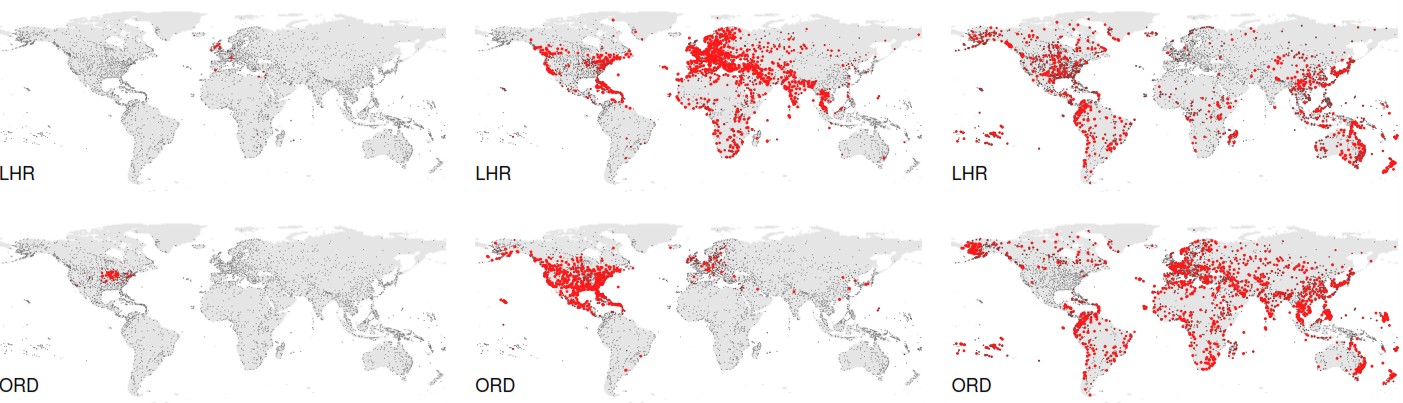}
\caption{Generic dynamic characteristics of global epidemic spread. Each panel depicts a temporal snapshot of a simulated pandemic that was generated by the metapopulation model of Eqs.~(\ref{eq:dynamicalsystem}). The top row depicts the time course for an initial outbreak in London, the bottom row in Chicago. From left to right the panels depict times $T=52,\,87$ and $122$ days since outbreak. Parameters of the simulation were $R_{0}=1.5$, $\beta=0.28$ and $\gamma=2.8\times10^{-4}.$ Because of the multi-scale nature of the mobility network and strong spatial heterogeneity the spreading patterns are spatially incoherent, complex and chaotic, no clear wave front exists from which a spreading speed could be computed, and arrival times at various places cannot be estimated on geometrical grounds.}
\label{fig:metapop}
\end{figure}

To best understand the approach, consider a metapopulation network, consisting of $n=1,...,M$ coupled populations, in which local disease dynamics are captured by SIR kinetics~\cite{anderson1991infectious}:
\begin{eqnarray}
I_{n}+S_{n} & \xrightarrow{\alpha} & 2I_{n}\label{eq:infect}\\
I_{n} & \xrightarrow{\beta} & R_{n}\label{eq:recover}\\
X_{m} & \xrightarrow{w_{nm}} & X_{n},\label{eq:move}
\end{eqnarray}
where $S_{n},I_{n},R_{n}$ denote the states (susceptible, infected, recovered) of an individual in population $n=1,....,M$. The first reaction represents disease transmission, the second recovery, and the third equation movement of a host from population $m$ to $n$ ($X$ is a placeholder for all three classes of individuals, i.e. $S,I$ and $R$). The parameters of this system are the population averaged effective per capita transmission rate $\alpha$, the population averaged recovery rate $\beta$, and the mobility rate $w_{nm}$. The mobility rate is given by the per capita traffic flux from $m$ to $n$, i.e. $w_{nm}=F_{nm}/N_{m}$, where $N_{m}$ is the size of population $m$. The basic reproduction number is defined according to
\begin{equation}
R_{0}=\alpha/\beta,\label{eq:R0}
\end{equation}
i.e. the expected number of secondary infections caused by one infected individual in an entirely susceptible population while it is infectious. The dynamics of the metapopulation SIR model, in the most parsimonious form, are given by
\begin{eqnarray}
\partial_{t}I_{n} & = & \alpha S_{n}I_{n}/N_{n}-\beta I_{n}+\sum_{m\neq n}\left(w_{nm}I_{m}-w_{mn}I_{n}\right)\nonumber \\
\partial_{t}S_{n} & = & -\alpha S_{n}I_{n}/N_{n}+\sum_{m\neq n}\left(w_{nm}S_{m}-w_{mn}S_{n}\right)\label{eq:SIRspatial}\\
\partial_{t}R_{n} & = & \beta I_{n}+\sum_{m\neq n}\left(w_{nm}R_{m}-w_{mn}R_{n}\right)\nonumber
\end{eqnarray}
(recycling the symbols $S,I,R$ for the number of individuals in a given state). Expressing the above in terms of fractions of susceptible, infected and recovered individuals ($s_{n}$, $j_{n}$, and $r_{n}$ respectively), the above dynamical system simplifies to
\begin{align}
\partial_{t}j_{n} & =\alpha\, s_{n}j_{n}-\beta j_{n}+\gamma\sum_{m\neq n}P_{mn}\left(j_{m}-j_{n}\right),\nonumber \\
\partial_{t}s_{n} & =-\alpha\, s_{n}j_{n}+\gamma\sum_{m\neq n}P_{mn}\left(s_{m}-s_{n}\right),\label{eq:dynamicalsystem}
\end{align}
with $r_{n}=1-s_{n}-j_{n}$, see Ref.~\cite{Brockmann_2013}. The rate parameter $\gamma$ is the average mobility rate, i.e. $\gamma=\mathcal{F}/\mathcal{N}$ where $\mathcal{F}=\sum_{n,m}F_{nm}$ is the total traffic flux in the network and $\mathcal{N}=\sum_{n}N_{n}$ the total population in the system. The matrix $\mathbf{P}$ with $0\le P_{mn}\le1$ quantifies the fraction of the travelers with destination $m$ emanating from node $n$, i.e.\ $P_{mn}=F_{mn}/F_{n}$, where $F_{n}=\sum_{m}F_{mn}$. If we now consider each node in the metapopulation network as the catchment area of each airport in the global mobility network and take $F_{nm}$ as the passenger flux between airports $n$ and $m$, the above dynamical system constitutes a structurally simple model for the global spread of an emergent infectious disease. In fact, the earliest successful models that incorporated worldwide mobility were structurally similar to Eqs.~(\ref{eq:dynamicalsystem}) and most sophisticated models (e.g the GLEAM simulation framework) possess an underlying mathematical machinery of similar nature~\cite{Hufnagel:2004kt, Colizza:2007p521, Colizza:2007p1066, grady2012}. Essentially the above model has three rate parameters, $\alpha,\beta$ and $\gamma$. Typical values for influenza-like disease are $\beta\approx0.3\,\text{days}^{-1}$, and $R_{0}=\alpha/\beta\approx1.3-2.3$. The global mobility parameter is typically much smaller $\gamma\approx0.001-0.02\,\text{days}^{-1}$. This implies that mobility has the slowest time-scale of the system. The complexity of the above model is encapsulated in the flux fraction matrix elements $P_{mn}$ which is shaped by the spatial heterogeneity of the metapopulation and the multiscale properties of the global mobility network. It is this coupling matrix that causes solutions of the above model to appear spatiotemporally complex, as is shown in Fig.~\ref{fig:metapop}. The figure depicts temporal snapshots of two simulated pandemics (i.e. solutions to Eqs.~(\ref{eq:dynamicalsystem})) with initial outbreaks in London and Chicago. Because of the inherent complexity in the global mobility network, i.e. the coupling matrix $P_{nm}$, the patterns appear to be complex, spatially incoherent and quickly lose any correlation to their initial outbreak location. Clearly, no geographic wave front is discernible except in the very beginning of the spreading process. The lack of geographic imprint makes it difficult to predict arrival time sequences from the maps alone. Furthermore, the geographic de-correlation makes it impossible to reconstruct the initial outbreak location based on a single temporal snapshot of the process. Unlike historic spreading processes that evolved according to more regular wave propagation, modern global disease dynamics make it difficult to define a proper propagation speed by, for instance, correlating geographic distance with arrival times.

\begin{figure}
\includegraphics[width=1\columnwidth]{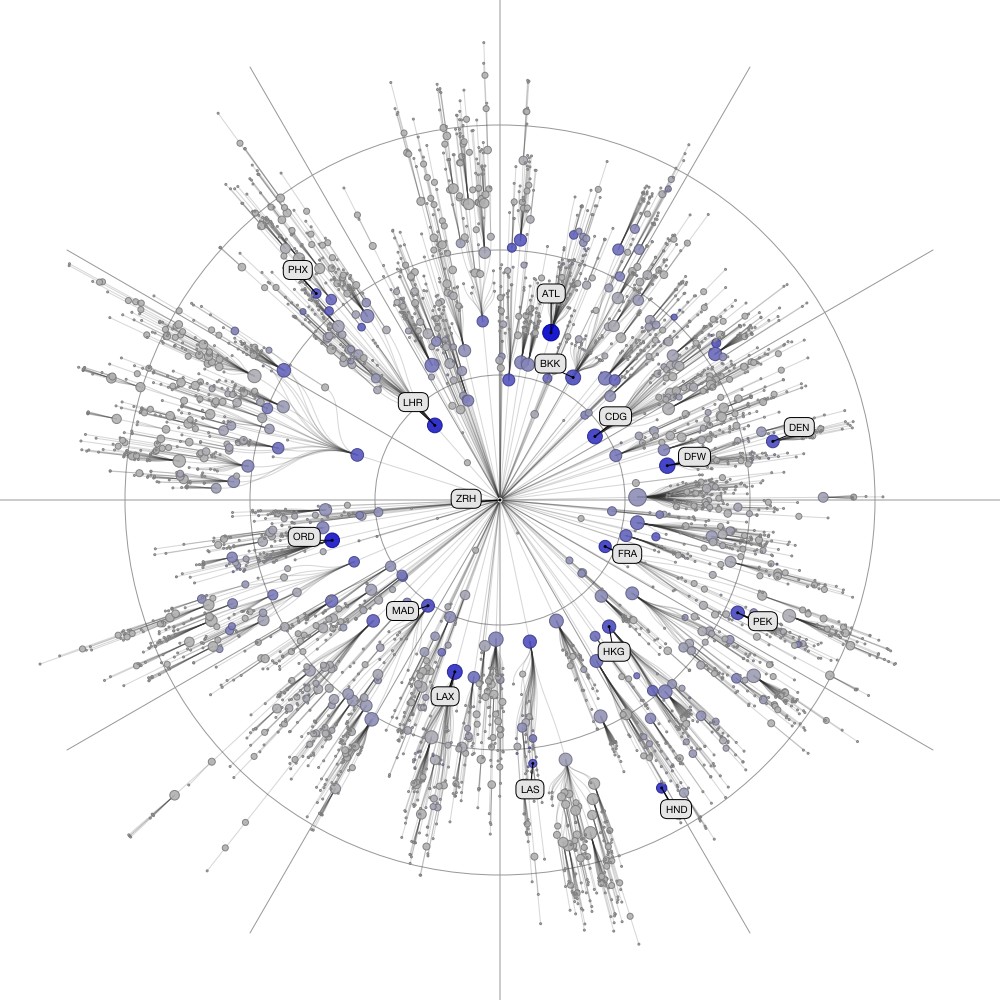}
\caption{Effective distance in the global mobility network: From the perspective of Zurich, each other airport has an effective distance derived from the entire worldwide air transportation network. In the illustration the effective distance is proportional to the radial distance from the root node ZRH. The underlying tree is the shortest path tree that reflects the most probable pathway of a contagion process proliferating through the network. Each symbol in the diagram represents one of the airports in the network. Symbol color quantifies the size of the airport and symbol size the number of branches at a node in the tree. A few major hubs in the network are labeled by their three letter airport code.}
\label{fig:spt}
\end{figure}

Because the spatio-temporal complexity generated by the model is caused by the coupling matrix $P_{nm}$, a plausible strategy for defining an effective distance measure must relate this distance to the coupling matrix. A possible choice is the following definition of an effective, directed length of a link $m\rightarrow n$ in the network:
\[
d_{nm}=1-\log P_{nm}>1
\]
If we interpret the flux fraction $P_{nm}$ as the conditional probability that an infection that is present at node $m$ will be passed to $n$, then the effective length from $m\rightarrow n$ is small if this probability is large, and vice versa. Now consider a multi-leg path $\Gamma=\{n_{1},\ldots,n_{L}\}$ through the network. Using the above definition, the effective, directed length of this path is the sum of the effective lengths of each of its legs. This implies that this length decreases as the probability increases that a path will be followed in a random walk process with hopping probabilities governed by matrix $P_{nm}$. Based on this, the effective distance $D_{nm}$ from an arbitrary reference node $n$ to another node $m$ in the network is defined by the length of the shortest path from $n$ to $m$:
\begin{equation}
D_{nm}=\min_{\Gamma}\lambda(\Gamma)\label{eq:effective_distance}
\end{equation}
This is equivalent to the assumption that the most probable path dominates, much like the smallest resistor in an electrical network with parallel conducting lines. From the perspective of a chosen node $n$, the set of shortest paths constitutes a shortest path tree $\tau_{n}$, corresponding to the most probable hierarchical sequence of pathways of an epidemic that originates at a reference node. Fig.~\ref{fig:spt} illustrates the properties of effective distance and the shortest path tree from sample reference node Zurich.

\begin{figure}
\includegraphics[width=1\columnwidth]{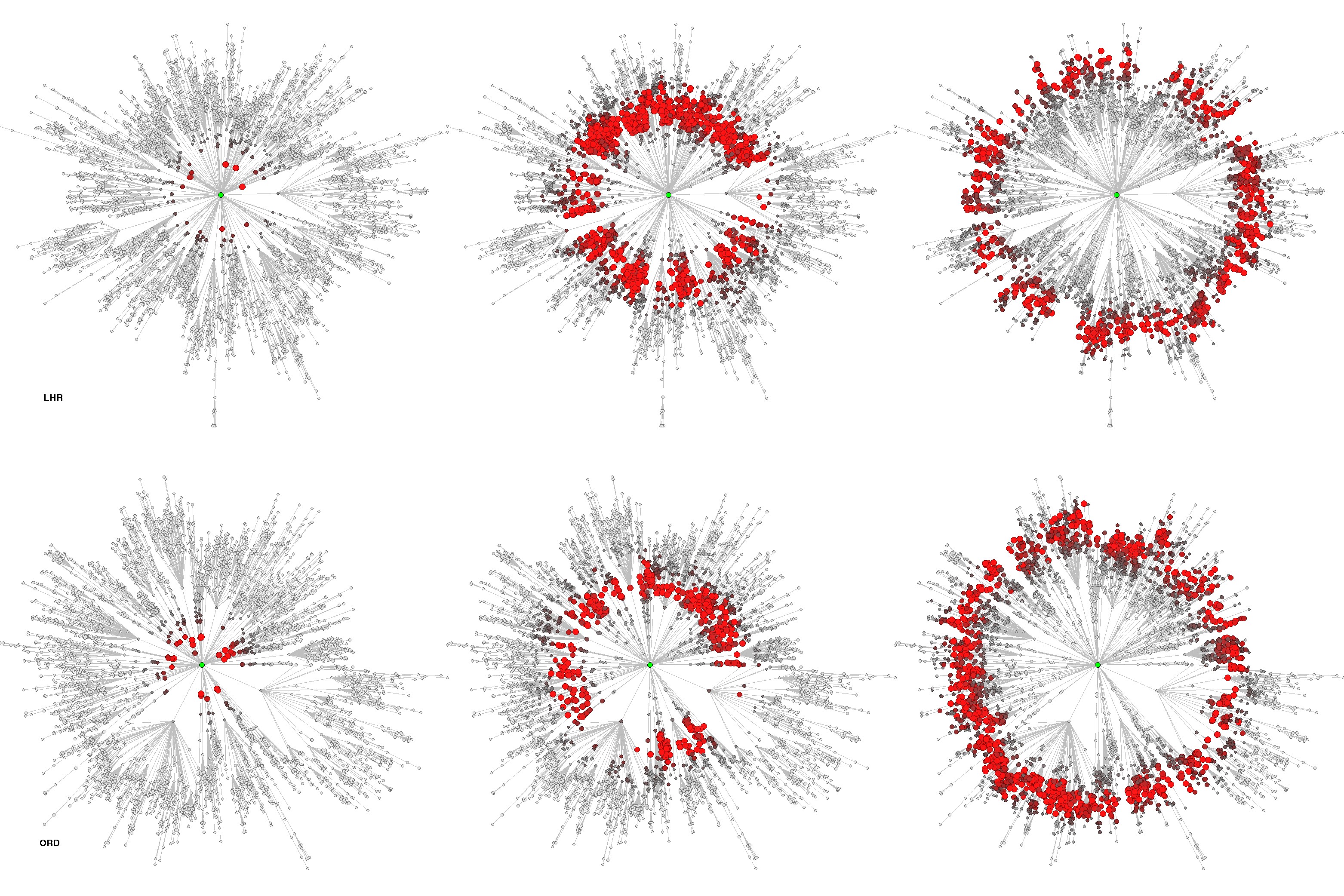}
\caption{Regular wave pattern in effective distance. The panels depict the same simulations as are shown in Fig.~\ref{fig:metapop} except from the effective distance perspective of the initial outbreak location, i.e. London (top row) and Chicago (bottom row). The pattern that appears to be complex in the conventional geographic view is mapped onto a regular wavefront pattern in the effective distance view.}
\label{fig:circular}
\end{figure}

The key question is: What are the properties of a simulated pandemic generated by the metapopulation model of Eqs.~(\ref{eq:dynamicalsystem}) in this new, effective representation, when viewed from the initial outbreak location? This is shown in Fig.~\ref{fig:circular}, which depicts the same simulation as is shown in Fig.~\ref{fig:metapop}, in the shortest path tree representation of each outbreak location.

We see that the patterns that appeared complicated in the conventional geographic representation exhibit regular wavefronts in the effective distance representation. Furthermore, these effective waves propagate at constant speeds. Note that this is more than just a simple remapping of a complex pattern onto a simple one. First, the observation of the propagating wavefront implies that the original assumption that contagion phenomena are dominated by most probable paths was correct. Furthermore, the remapping only depends on the coupling matrix $P_{nm}$ in the original nonlinear dynamical system. Because of this, one expects to observe concentric waves irrespective of the other parameters of the model, e.g. the rate parameters $\alpha$, $\beta$ and $\gamma$.
\par
The advantages of the effective distance approach become immediately apparent. Because regular wave propagation is observed in the effective distance view, one can easily compute arrival times at every node in the network if the effective speed of the wavefront is known. Also, because the effective distance only depends on the topological features of the underlying network, we can infer an approximate factorization of the parameters that determine the arrival time of the epidemic wavefront at any node in the network:
\begin{equation}
T_{a}=\underbrace{D_{eff}(\mathbf{P})}_{\text{ eff. distance}}/\underbrace{v_{eff}(\beta,R_{0},\gamma)}_{\text{ eff. speed}}.
\end{equation}
The arrival time can be computed by the ratio of effective distance from the outbreak origin, which only depends on matrix elements $P_{nm}$, and the effective speed, which in turn depends on the rate parameters of the dynamical systems. This implies that, even if these rate parameters are unknown, one can still estimate the relative arrival times or the sequence of arrivals of an epidemic, irrespective of disease specific parameters. For example, given an outbreak at node $n_{0}$ and denoting the arrival time at any other node $m$ by $T_{a}(m|n)$ and the effective distance from $n_{0}$ to $m$ by $D_{eff}(m|n_{0})$ we have:
\[
\frac{T_{a}(m|n_{0})}{T_{a}(n|n_{0})}=\frac{D_{eff}(m|n_{0})}{D_{eff}(n|n_{0})}.
\]
When new pathogens emerge, for instance the recent MERS-CoV (Middle East Respiratory Syndrome - Coronavirus), it is typically unknown whether the new agent has the potential to develop into a full-scale pandemic because: (1) careful measurement of the relevant disease specific parameters is time consuming, (2) typically transmission pathways are still unknown and must be unravelled, and (3) recovery rates and other properties and dependencies must me assessed. The effective distance approach shows that complexity science, and network science in particular, can help extract properties of a potential spread that are largely robust against variations in these parameters. Thus universal statements about a potential global spread can be made. This is very valuable information during the onset of an epidemic since one can initiate effective containment strategies with minimal information concerning the disease. Because the approach is so general and is entirely based on the use of a more appropriate notion of distance in heterogeneous network systems, it is clearly applicable to other types of contagion processes that occur on networks, such as the spread of news, information or fads in social networks.
\par
Another insight gained by the approach is that one observes concentric wave patterns only from the perspective of the outbreak origin. From the perspective of any other node in the network, the dynamic pattern appears more irregular. This can be quantified and used to reconstruct the outbreak origin for contagion phenomena that have been spreading but are unnoticed until the incidence passes a certain threshold~\cite{Brockmann:2003tn}. Typically these incidence patterns bear no geographic regularity and it is difficult, albeit essential, to reconstruct the spatial origin in order to develop efficient containment strategies. The effective distance approach can help reconstruct the outbreak origin by identifying the node from which the observed pattern appears to be most regular. This approach has successfully reconstructed outbreak origins of recent epidemics, e.g. the foodborne disease EHEC/HUS that struck Germany and its European neighbors in 2011.

\subsection{Enabling more effective disease control using social information systems}
Vaccine-preventable diseases can be eradicated from a population when enough individuals immunize. Unfortunately, enforcing compulsory vaccination is challenging and potentially incompatible with personal liberties~\cite{colgrove2006state,verweij2004ethical,salmon1999health}. Beyond ethical and political concerns, the efficacy of compulsory vaccination programs may be constrained by (1) the limited information available to those designing the policies, and (2) uncertainty about how people will change their behavior~\cite{funk2010modelling} in response to the information they receive as an epidemic progresses. Here we will investigate when \emph{voluntary} vaccination~\cite{bauch2004vaccination,bicknell2002case} can be a viable alternative to imposing immunization in a top-down way. That is, we use a model to determine if and when information about the risk of becoming ill will enable individuals to voluntarily make decisions that are beneficial, not only to their own health, but also to the health of many others.

\begin{figure}[htbp]
	\centering
	\includegraphics[width=1\textwidth]{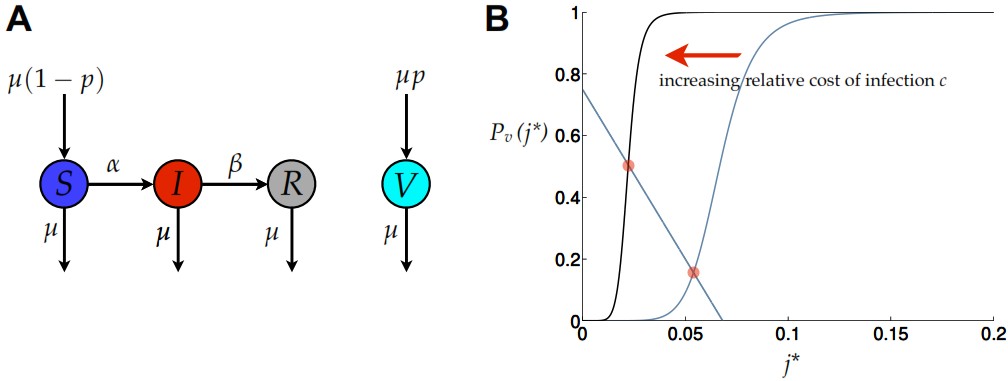}
	\caption{The vaccination model in a well-mixed population. (A) The individuals in the population can be in one of four states: Susceptible, Infected, Recovered or Vaccinated. Individuals enter/exit the population at the birth/death rate $\mu$. At birth individuals are susceptible unless they vaccinate, which happens with probability $p$. Susceptibles become infected through contact with infecteds at contact rate $\alpha$. Infecteds recover with recovery rate $\beta$. (B) The intersection of the $j^*$ and $P_v(j^*)$ curves gives the steady-state infection in the system. As $c=c_i/c_v$ increases, $j^*$ decreases, but remains non-zero for finite $c$, indicating that disease cannot be eradicated for $c_i$ finite and $c_v>0$.}
	\label{fig:vaccination1}
\end{figure}

To model vaccination we consider the SIR model (see Sec.~\ref{sec:epidemics1}) with demography and assume that a fraction $p$ of individuals vaccinates at birth (see Fig.~\ref{fig:vaccination1}A). Commonly, $p$ is taken to be a constant. In a population where contact occurs at random between any pair of individuals (known as the well-mixed assumption) the dynamics of this model are described by the following system of ordinary differential equations (which are exact in the limit of infinite population size $N$):
\begin{align}
\qquad\partial_{t}{s} &= -\alpha sj-\mu s+\mu (1-p) \nonumber\\
\partial_{t}{j} &= \alpha sj- \beta j- \mu j \label{eq:pop_level}&\\
\partial_{t}{r} &= \beta j - \mu r \nonumber\\
\partial_{t}{v} &= -\mu v + \mu p \, , \nonumber
\end{align}
where $s$, $j$, $r$ and $v$ are the fraction of individuals that are respectively susceptible, infected, recovered and vaccinated. We assume that the population size is approximately constant, and therefore the natural birth and death rate are both given by $\mu$. The contact rate between individuals is $\alpha$, and  $\beta$ the recovery rate. (The fractions denoting the state of the system are a function of time although we do not explicitly show this dependence.) Setting the time derivatives to zero we can easily determine the steady-state infection level $j^*$ as a function of $p$:
\begin{equation}
	j^{*}=\frac{\mu(1-p)}{\beta+\mu}-\frac{\mu}{\alpha} \, . \label{eq:equi_infection}
\end{equation}
This implies that the disease will die out when the vaccinated fraction of individuals is $p>1-(\gamma +\mu)/\alpha$ (see e.g.~\cite{keeling2011modeling} for a more detailed derivation).
\par
To model voluntary vaccination we modify this standard model so that the fraction of individuals who vaccinate is a function of the fraction of infected individuals rather than a constant. This captures the idea that individuals will perceive a greater risk of becoming ill when they see higher disease incidence around them, and are therefore more likely to vaccinate~\cite{funk2010modelling}. More specifically, following prior models~\cite{bauch2004vaccination} and experimental evidence~\cite{chapman2012using}, we assume that individuals make rational choices. In this framework an individual who vaccinates will incur a cost $c_v$ (the cost of vaccination) and one that becomes ill will incur the cost of infection $c_i$ (we assume that vaccines have $100\%$ efficacy and thus vaccinated individuals will never become infected, but this assumption can be easily relaxed). Therefore, the expected cost to an individual who vaccinates is $c_v$. Meanwhile, the expected cost to an individual who does not vaccinate is $c_i R$,
where $R$ is an individual's estimate of the probability that she will fall ill if she does not vaccinate. We further assume that $R$ is simply estimated to be the fraction of infected others she sees; thus $R=j$ in a well-mixed model.
\par
A rational individual will choose to vaccinate if this action minimizes the cost she expects to incur. Let $P_v$ be a Boolean variable that indicates whether an individual vaccinates ($P_v=1$ denotes vaccination). Then an individual's choice will follow the threshold rule:
\begin{equation}	
 P_v = \begin{cases}
    1 & \quad\text{if $c R \leq 1$} \\
    0 & \quad\text{if $c R > 1$} \, , \label{eq:choose_fun}
\end{cases}
\end{equation}	
where $c=c_i/c_v$ is the perceived relative cost of infection.
We will approximate this step function by the continuous function
\begin{equation}
	P_v(R;c,\omega)=\frac{(c R)^{\omega}}{(c R)^{\omega}+1} \, . \label{eq:vac_choice}
\end{equation}	
This function is more general than the sharp threshold in Eq.~(\ref{eq:choose_fun}) because the smoothness incorporates small variation in the vaccination threshold of a single individual (due to e.g. individual error) and also variation across the thresholds of different individuals. (In the limit $\omega\rightarrow\infty$ we recover the discontinuous threshold rule; here we fix $\omega$ at the intermediate value $\omega=8$.)
\par
To approximate the well-mixed dynamics in a system where individuals make vaccination choices according to rule (\ref{eq:choose_fun}), we use Eq.~(\ref{eq:pop_level}) with $p=P_v$. Thus, the steady-state infection level $j^*$ is given by Eq.~(\ref{eq:equi_infection}), represented graphically by the intersection of the curves in Fig.~\ref{fig:vaccination1}B. The crucial insight to draw from this solution is that, as long as there is any cost associated with vaccination, there will be insufficient vaccination to eradicate a disease. A similar result, echoing the troubling conclusion that voluntary vaccination is not a viable alternative to enforced immunization, was derived through a game-theoretic argument in \cite{bauch2004vaccination}.
\par
\begin{figure}[h]
\centering
\includegraphics[width=1\textwidth]{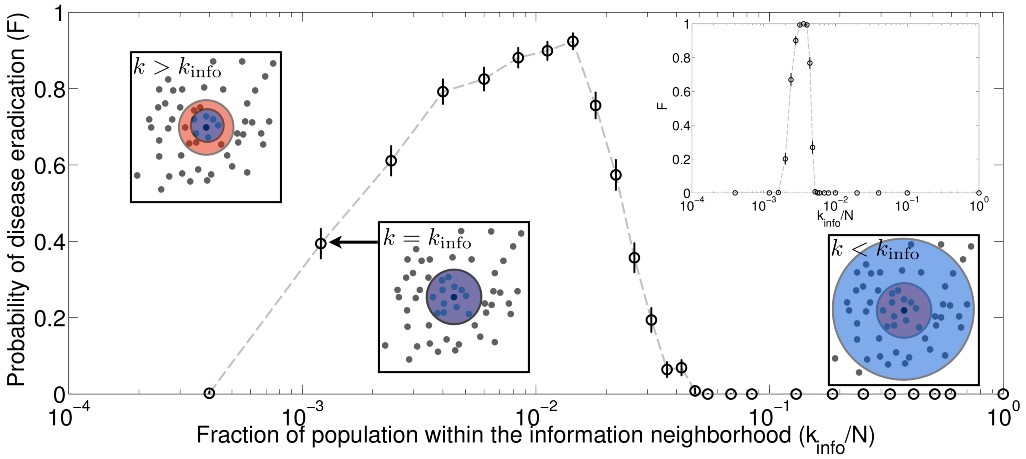}
\caption{The right information range facilitates disease eradication. Probability $F$ that the disease is eradicated after a fixed time $T$, as a function of $k_{\text{info}}/ N$, the fraction of the total population within an individual's information neighborhood (here $N=10^4$ agents). The main figure is for a lattice network while the inset shows a random network, both with a contact neighborhood of size $k=12$. An individual interacts with others in her contact neighborhood (red) and sees the state of those in her information neighborhood (blue). The overlapping area is shown in violet. The information neighborhood is a subset of the contact neighborhood when $k>k_{\text{info}}$, exactly the same when $k=k_{\text{info}}$, and a superset when $k<k_{\text{info}}$.}
\label{fig:vaccination2}
\end{figure}

The prior result relies on the assumption that the spread of disease, and the information available to individuals about this spread, are highly homogeneous throughout the population. However, as discussed in Sec.~\ref{sec:epidemics1}, the complex contact patterns between individuals, which mediate the spread of diseases in the real world, lead to spatio-temporal variation in infection levels. Therefore, we ask how the effectiveness of voluntary vaccination changes when individuals make immunization choices using information that incorporates this variability. To answer this question we introduce a model where individuals interact physically only with a subset of the population, which we call \emph{contact neighbors}. We employ the formalism of a network~\cite{newman2009networks} to represent this contact structure~\cite{keeling2005networks, pastor2002epidemic, newman2002spread}. Each individual is represented by a node, and the links between nodes represent the potential for an infectious interaction. To model variation in the information that individuals have about the health of others, we introduce a second network where individuals are connected if they are informed about each other's state. (The transmission of information could be through any mechanism: e.g. through direct conversation, word of mouth, observation of symptoms, mass-media reporting, or social media.) The information neighborhood can be (1) a subset of the contact network, (2) exactly equal to it, or (3) a superset (see Fig.~\ref{fig:vaccination2}). By construction, all individuals have the same number of contact neighbors $k$ and the same number of information neighbors $k_{\text{info}}$. However, we allow for the number of contact neighbors to differ from the number of information neighbors. The key point is that, in this setting, an individual estimates the risk of infection $R$ from the fraction of its infected \emph{information} neighbors (rather than considering the global fraction of infected individuals as above). For both of these networks we consider two types of topologies: (1) a network where every node has the same predetermined number of neighbors but otherwise has a random structure, and (2) a lattice network (for description of these two types of networks see a standard network reference, e.g. Ref.~\cite{newman2009networks}). These two simple network topologies are different enough to demonstrate that the results we find are not due to specific structural features.

To investigate the effect that the information \emph{range} has on the effectiveness of voluntary vaccination, we simulate an individual-level model of this system and vary the size of the information neighborhood $k_{\text{info}}$. The contact neighborhood $k$ is small and fixed (we set $k=12$, but the behavior is similar for neighborhoods up to an order larger). To estimate the probability that vaccination will remove the disease from the population, we fix a time $T$ much larger than the transient time, and count the number of simulations where $j(T)=0$. We find that this fraction $F$ of simulations where the disease is eradicated changes systematically with $k_{\text{info}}$ (see Fig.~\ref{fig:vaccination1}). Interestingly, $F$ exhibits a clear maximum when the information neighborhood is of an intermediate size. Access to information about a growing number of neighbors initially facilitates eradication of the disease. However, as individuals average over information obtained from increasingly large regions of the system, the disease becomes decreasingly likely to die out. Further experiments indicate that this qualitative behavior holds across different epidemiological parameters that allow the infection to invade the population when there is no vaccination\footnote{The condition for a disease to invade the population  is that $\alpha/(\beta+\mu) > 1$~\cite{Kermack:1927p414}. This is one of the most established results in mathematical epidemiology.}, although the mechanisms driving the behavior vary~\cite{VaccinationPaper}.
\par
In summary, although an epidemic cannot be contained by voluntary immunization when individuals use average global information, the successful eradication of a disease can occur for an intermediate range of information that is neither local nor global. In this range individuals make their vaccination decisions taking into account variability in infection densities. This suggests a promising alternative to top-down vaccination policies: empowerment of people through information systems such as social media, thereby allowing them to take responsible decisions to immunize themselves. Importantly, given the right information range, individual choices will be aligned with socially desirable outcomes.

\section{Summary, conclusions and outlook}
In this paper, we have discussed recent work on crowd disasters, crime, terrorism, war, and the spread of disease from the perspective of complexity science. We have pointed out that many counter-intuitive system behaviors are due to the existence of non-linear amplification, feedback, and cascade effects. Many of the surprising effects discussed here result from correlations between dynamical processes that invalidate the common representative agent or mean-field approach, which is based on the assumption of well-mixed interactions. This has important policy implications, namely that conventional approaches to address the above problems can potentially deteriorate the situation, while their mitigation requires a different perspective and a new approach. Therefore, old problems can be addressed more successfully when applying a complex systems perspective.
\par
We have pointed out that the classical rational choice approach suggests that problems such as crime or terrorism should be effectively suppressed by strong enough deterrence. In contrast to these expectations, however, one finds crime cycles, escalation and recurrence of conflicts. The reason for this is that the equilibrium or representative agent approaches underlying classical rational choice analysis are not always applicable~\cite{NATURE}. The social systems investigated here are characterized by systemic instabilities. Under such conditions, the desirable system state is left sooner or later, often giving rise to cascade effects such as contagious spatio-temporal spreading.\footnote{Bankruptcy cascades~\cite{bankruptcy} are a good example for this.} As a result, the system is eventually driven out of control, despite everybody's best intentions and efforts to avoid this.
\par
Today's prevalent focus on individuals, their behavior and how to control it, ignores the importance of systemic effects. For example, recent reports have cast serious doubt on the effectiveness of deterrence strategies.\footnote{A recent report on CIA practices calls the effectiveness of ``harsh interrogation'' techniques into question (http://www.huffingtonpost.com/2014/03/31/senate-torture-report\_n\_5061921.html, accessed Apr. 16, 2014), and an independent US panel recently concluded that mass surveillance of phone data has been ineffective in detecting terrorist threats (http://www.bbc.co.uk/news/world-us-canada-25856570, accessed Jan. 24, 2014). Recent violent clashes in Rio de Janeiro, Brazil, have cast doubt on the effectiveness of massive police interventions in pacifying the city's favelas (http://www.theguardian.com/world/2014/apr/23/rio-deadly-clashes-death-of-dancer, accessed April 24, 2014).} Calls for more or longer prison sentences and extended surveillance stress that current punitive approaches are not working well. Despite the best intentions, ``more of the same'' is very unlikely to have positive effects if we do not understand why such measures have been of limited effect in the past.
One of the key problems in this regard appears to be the individual-centric approach, which often systematically neglects the collective dynamics, i.e. the complex systemic interplay of individual actions and the broader social context they take place in as well as the socio-economic conditions of an individual.
\par
More successful approaches for averting system dynamics that endanger human lives need to move from an individual-centered to a systemic perspective. For example, harmful cascade effects resulting from systemic instabilities and related loss of control may be countered by suitable system designs with engineered breaking points or adaptive decoupling strategies such as separation or immunization~\cite{NATURE}:
\begin{enumerate}
\item For example, when the density is so high that there are involuntary body contacts between many people, crowd disasters are quite likely. Under such conditions, forces are transferred between the bodies of people. They may push them around in a phenomenon called ``crowd turbulence'', such that some individuals may stumble. If someone is falling to the ground, a domino effect is likely to cause many people to fall over each other. As a consequence, the people on the ground may suffocate. To avoid this, evacuation (pressure relief strategies) are required early on, to avoid densities under which forces may be transmitted between bodies. In other words, low enough densities will avoid cascade effects. One possible way of reaching this is the use of {\it apps} that provide re-routing recommendations to visitors of mass events.
\item We have also seen that crime may spread by imitation of successful neighbors. In our model, we have studied the effect of peer punishment (which also underlies the establishment of social norms). Surprisingly, crime cannot be eliminated by more deterrence and high discovery rates. The reason is that (most) people are not criminal by nature. Their behavior rather depends on the behaviors of their neighbors in the social interaction network. Hence, improving socio-economic conditions may be more effective than sending many people to prison.
\item Furthermore, deterrence may not only fail to stop terrorism but, on the contrary, may lead to more attacks. We showed that unspecific interventions causing harm to civilians tend to undermine the legitimacy of the ``war on terror'', which may instead increase civilian support for the cause of ``freedom fighters''. This, in turn, can ultimately lead to escalation instead of the intended de-escalation of violence.
\item In the context of the conflict in Jerusalem we discussed the effect of socio-cultural distance between different population groups on levels of violence. Deterrence or policing alone does not appear to be sufficient to deter violence between the various factions in the city. If the socio-cultural distance between the groups is too large, encounters in daily life may spark violent confrontations. Given large socio-cultural distances that can not easily be changed, an alternative and more effective strategy may be to limit the contact between groups and, thus, reduce the number of possible triggers and interrupt cascade effects.
\item A complementary approach to reducing violence is to predict the onset of conflict before it actually occurs. We discussed a methodology based on the analysis of large news archives to detect growing tensions that are indicative of potential escalations. In line with findings in complex biological and natural systems, early warnings signals for critical transitions can be found not only by direct measurement of the phenomenon---an increase in the number of conflict-related news prior to conflict---but also by other attributes of the time series, such as changes in variance over time. Based on such signals, international political, social, or economic efforts may be made to reduce upcoming tensions and thereby thwart confrontations and war.
\item Cascade effects do not only promote crowd disasters, crime and violence. They are also the basis of the epidemic spreading of diseases. We have presented an effective distance measure that can serve to understand the spatio-temporal infection dynamics, to identify the origin location of a disease, and to predict its further spreading. Thereby, the model can help to use limited immunization doses in a better way to contain pandemics more effectively. In order to reach the best possible immunization, well-adapted decentralized information strategies, e.g. through social media, promise to be most effective.
\end{enumerate}
The above insights can contribute to saving human lives by promoting approaches that differ markedly from commonly applied strategies. In particular, the findings discussed in our paper suggest that too strict control attempts may actually cause a loss of control in a complex system. The reason is that strong control may not only be expensive, but also ineffective, as it tends to undermine the natural capability of complex systems to self-organize.\footnote{In some sense, strong control may inhibit the normal functionality of a complex system. An example for the failure of top-down control is the fact that even the most sophisticated technological control mechanisms increase flight safety less efficiently than a non-hierarchical culture of collaboration in the cockpit where co-pilots are encouraged to question the decisions and actions of the pilot \cite{airsafety}. Moreover, the official report on the Fukushima disaster \cite{nuclearsafety} stresses that it was not primarily the earthquake and tsunami which was responsible for the nuclear meltdowns, but ``our reflexive obedience; our reluctance to question authority; our devotion to `sticking with the program'; our groupism.''---in other words, too much top-down control. Similar problems are likely to arise from the application of mass surveillance, as it is likely to cause self-censorship and to reduce socio-economic diversity, thereby decreasing the capacity of a society to innovate and adapt quickly enough to our changing reality (e.g. to the rapid digital revolution). Therefore, security is not just an issue of suppressing malicious behavior. In the first place, it requires a resilient and adaptive society, which allows and empowers people to control their own lives.} To control a complex system, a ``long leash'', i.e. allowing for some flexibility (``freedom'') often works better than a ``short leash''.\footnote{The American prohibition, which actually led to an increase in violent crime~\cite{Miron_1999}, is probably a good example for unintended detrimental effects of strict regulations.} A promising approach along these lines is the principle of ``guided self-organization''~\cite{NATURE, ManComp}, i.e. to interfere with the system in a minimally invasive way, such that the system self-organizes and does by itself what we want it to do. In this context, it is worth reminding the reader of the ``minimally invasive'' principle of ``chaos control'', which just pushes the system slightly, at the right moment, to let it take a particular path~\cite{ChaosControl}.
\par
Planning and control approaches mostly work well for stationary systems, while complex systems with extremely variable and hardly predictable dynamics call for enough autonomy to flexibly adapt to the respective local requirements~\cite{autonom}. In other words: managing complexity requires decentralized adaptation rather than centralized control. Nevertheless, it is possible (and even necessary) to enable the self-organization of complex systems~\cite{socialselforganization}. Suitable institutional settings, such as proper organizational frameworks for mass events, can support systems in their self-organization and self-regulation, thereby reducing the need for interference. This means replacing (or complementing) the classical concept of planning and top down control with the concept of empowerment, i.e. creating opportunities for individuals to make desirable decisions rather than to enforce a particular way of reaching a goal~\cite{enabling, economy}. In fact, to be manageable, complex systems must be designed in such a way that the interactions and institutional framework jointly enable favorable self-organization.

\subsection*{Acknowledgments}
This work has benefitted from the FET Flagship Pilot Project FuturICT (grant number 284709), from the ERC Advanced Investigator Grant ``Momentum'' (grant number 324247),
and the ETH project ``Systemic Risks, Systemic Solutions'' (CHIRP II project ETH 48 12-1). MP acknowledges support from the Slovenian Research Agency (Grant J1-4055 and Program P5-0027).

\subsection*{Author contributions}
DH has set up the author team and scientific concept of this paper. All authors have contributed to the writing of this manuscript. MM, AJ, DH contributed research on crowd turbulence, MM, JK, DH on simple rules for safety, UB on apps for saving lives, KD, MP and DH on crime, KD and SS on insurgent conflict and terrorism, TC on interstate wars, and DB, DH, OWM on disease spreading.

\end{document}